\documentclass[twocolumn,showpacs,superscriptaddress,prb,amsmath,amssymb,letterpaper]{revtex4}
\usepackage{times}
\usepackage{amsmath,bm,amsfonts}
\usepackage{dcolumn}
\usepackage{graphicx}
\usepackage{latexsym}

\begin{document}

\title{Skyrmion quantum numbers and quantized pumping in two dimensional topological chiral magnets}

\author{Bohm-Jung \surname{Yang}}
\affiliation{Correlated Electron Research Group (CERG),
RIKEN Advanced Science Institute, Wako, Saitama 351-0198, Japan}

\author{Naoto \surname{Nagaosa}}
\affiliation{Correlated Electron Research Group (CERG),
RIKEN Advanced Science Institute, Wako, Saitama 351-0198, Japan}

\affiliation{Department of Applied Physics, University of Tokyo,
Tokyo 113-8656, Japan}

\affiliation{Cross-Correlated Material Research Group (CMRG),
RIKEN Advanced Science Institute, Wako, Saitama 351-0198, Japan}

\date{\today}

\begin{abstract}
We investigate the general conditions to achieve the adiabatic charge and spin
polarizations and quantized pumping
in 2D magnetic insulators possessing inhomogeneous spin structures.
In particular, we focus on the chiral ferrimagnetic insulators which are generated via spontaneous symmetry breaking from
correlated two dimensional topological insulators.
Adiabatic deformation of the inhomogeneous spin structure generates the spin gauge flux,
which induces adiabatic charge and spin polarization currents.
The unit pumped charge/spin are determined by the product of two topological invariants
which are defined in momentum and real spaces, respectively.
The same topological invariants determine the charge and spin quantum numbers of skyrmion textures.
It is found that in noncentrosymmetric systems, a new topological phase, dubbed the topological chiral magnetic insulator, exists
in which a skyrmion defect is a spin-1/2 fermion with electric charge $e$.
Considering the adiabatic current responses of generic inhomogeneous systems,
it is shown that the quantized topological response of chiral magnetic insulators
is endowed with the second Chern number.
\end{abstract}

\pacs{}

\maketitle

\section{\label{sec:intro} Introduction}
After the seminal papers written by Jackiw and Rebbi~\cite{Jackiw}
and Su, Schrieffer, and Heeger~\cite{Su},
quantum number fractionalization has become one of
the most fundamental and fascinating concepts in condensed matter physics.
For one dimensional spinless fermions, a domain wall soliton
induces a localized mid-gap state which has a half of the electric charge.
Later this idea is further generalized by Goldstone and Wilczek.~\cite{Goldstone}
Considering slow variations of the background mass fields in space and time,
adiabatically induced currents are derived.
The time component of the adiabatic current describes the charge density
induced by the spatial variation of the mass soliton.
The spatial integration of this charge density gives rise to the half electric charge.~\cite{Goldstone}
Moreover, the temporal variation of the background mass fields
generates the spatial component of the adiabatic current,
which describes the adiabatic charge polarization
or even the quantized charge pumping.~\cite{Qi}
Therefore the existence of the topological soliton with nontrivial quantum numbers
reflects the quantized dynamical response of
the system.

The discovery of the time-reversal invariant spin Hall insulators~\cite{Murakami} has
triggered great attentions to the
two dimensional topological band insulators (TBIs).
TBIs are the insulating phases that carry
gapless edge states at the sample boundary due
to the nontrivial bulk Berry phase,  which
can either break~\cite{Haldane} or preserve the time reversal symmetry.~\cite{SHI}
Because of the distinct bulk topological properties,
$\pi$ flux defects in TBIs carry nonzero spin or charge quantum numbers
leading to Jackiw-Rebbi type topological solitons in two dimensional systems.~\cite{DHLeesoliton}
In addition, in strongly interacting electron systems, it is proposed that TBIs can be obtained
from the spontaneous breaking of spin rotation symmetry.~\cite{TMI} Since this topological ``Mott" insulator
possesses a vector order parameter $\hat{N}$,
skyrmion-type topological solitons are allowed,
which are also expected to have nontrivial quantum numbers.
For example, in the case of the spin Hall ``Mott" insulator,
a skyrmion defect is a boson carrying a quantized electric charge 2$e$.~\cite{Senthil}

In this work, we focus on a different class of topological Mott insulators,
the chiral magnetic insulators (CMIs),
which can be obtained by considering
electron-electron interactions in TBIs.
In CMIs, the vector order parameter $\hat{N}$ is given by
the stagggered spin ordering, which can acquire
explicit time dependence by coupling to
the external magnetic field.
Because of the quantized
charge and spin Hall effects, a skyrmion defect in CMIs
can carry charge and spin quantum numbers satisfying fermionic or bosonic statistics
depending on the magnitude of the bulk topological invariants.
Considering smooth variations of the spin ordering directions in space and time,
reminiscent of the approach pursued by Goldstone and Wilczek,
we have shown that the quantized charge and spin pumpings are possible under
the adiabatic change of the inhomogeneous spin structures such as
the domain wall with rotating spins in collinear ferrimagnets or the
conical ferrimagnetic phases in which the spatial modulation
of the spin direction spreads over the whole system.

In CMIs, the quantum of the pumped charge/spin or the quantum number of
a skyrmion strongly depends on the nature of the parent
TBIs. In the case of the CMIs generated from
the anomalous Hall insulator (AHI),
the quantum of the pumped charge/spin is solely determined
by the spin/charge Chern numbers of the collinear magnetic ground states.
When the magnitude of the spin order parameter is small, the charge and spin
Chern numbers are the same as those of AHI. In this case,
the spin can be pumped in the unit of $\hbar$ while the charge pumping is forbidden,
which, at the same time, means that the skyrmion is a chargeless spin-1 boson.
As the magnitude of the ordered spin moment increases, a new CMI phase dubbed
the topological chiral magnetic insulator (TCMI) emerges in noncentrosymmetric systems, which supports
a fermionic skyrmion with charge $e$ and spin $\frac{\hbar}{2}$.
Here the quantized charge and spin pumpings can be achieved.

On the other hand, in the case of CMIs derived from the spin Hall insulator (SHI),
the spin anisotropy inherent to SHI due to the spin-orbit coupling imposes constraints
on the adiabatic pumping and skyrmion quantum numbers.
The polarizations strongly depend on the relative orientation of the spin anisotropy
and the magnetic ordering directions.
When the spin order parameter is parallel to
the spin anisotropy direction, the charge polarization
induced by the rotating spins in domain walls is always zero.
On the other hand, if the spin ordering is perpendicular to
the spin anisotropy direction, finite charge polarization
is expected for the Neel domain walls and the transverse
conical spin states.
In the case of a skyrmion, it does not carry a charge quantum number
independent of the magnitude of the spin order parameter.
However, a meron defect can be charged but the net charge is not quantized. 


In addition, it is shown that the charge and spin polarizations in CMIs
can be understood in the framework of the adiabatic
topological responses of generic inhomogeneous crystals.~\cite{Niu,Niu_review,Essin}
The inhomogeneity induced polarization current
contains a topological part whose topological nature is endowed with the second Chern number,
which is quantized in four dimensional closed manifolds.
When the inhomogeneity is introduced
by the spatially varying three-component unit vector $\hat{N}$,
the inhomogeneity induced topological current can be separated into the homogeneous and
inhomogeneous parts, which allows the second Chern number to be
represented by the product of two independent topological
invariants. One is the Chern number that
is quantized in the momentum space
and the other is the skyrmion number defined in the
position/time spaces.

Since electron correlation effects are required
to achieve CMI in addition to large spin-orbit coupling,
transition metal oxides with 5d electrons
can be an ideal platform to realize CMI.
Recently, it is confirmed that the ground state of Sr$_{2}$IrO$_{4}$
is a spin-orbit entangled Mott insulator with anti-ferromagnetic spin ordering,~\cite{bjkim1,bjkim2,sjmoon,hosub}
which manifests the strong interplay of electron correlation and spin-orbit coupling effects
in Ir-based compounds.
In the case of Na$_{2}$IrO$_{3}$ which is expected to be SHI in non-interacting limit,
anti-ferromagnetic spin ordering develops when electron correlation effects are included
on the parent SHI phase.~\cite{Shitade}
In this respect, Na$_{2}$IrO$_{3}$ is a candidate material in which
the topological response of  CMI can be observed.
Moreover, it is recently proposed that
bilayers of the perovskite-type transition metal oxides grown along the [111] direction
can be potential candidates for two-dimensional topological insulators.~\cite{111_1,111_2}
The transition metal ions in the bilayer are located on a honeycomb lattice
consisting of two trigonal sublattices on different layers.
Interestingly, since the layer potential difference can be easily created by applying an electric field
or by sandwiching the bilayer between two different substrates,
lattice inversion symmetry can be explicitly broken in this system.
Therefore to reveal the ground state of the bilayer systems,
it is important to understand the interplay of
the spin-orbit coupling, electron correlation, and the broken inversion symmetry.

The rest of the paper is organized in the following way.
In Sec.~\ref{sec:Model} we construct
a model Hamiltonian of CMIs taking into account
electron-electron interactions in TBIs on the honeycomb lattice.
Sec.~\ref{sec:TCMI_AHI} we consider the general conditions
to achieve the quantized charge and spin pumpings
through a domain wall for CMI derived from AHI.
The charge and spin quantum numbers of a skyrmion in CMI
are also discussed.
The investigation of the inhomogeneity induced charge polarization
and skyrmion/meron quantum numbers is extended to CMI dervied from SHI
in Sec.~\ref{sec:TCMI_SHI}.
In Sec.~\ref{sec:Semiclassical} the adiabatic polarization currents
for generic inhomogeneous systems are derived by using the semi-classical
gradient expansion method. The topological origin
of the quantized charge pumping in CMI is discussed from this viewpoint.
Finally, we conclude in Sec.~\ref{sec:Conclusion}.
The details of the procedures to derive the adiabatic polarization
currents from the gradient expansion method are presented in Appendix~\ref{sec:GradientExpansion}.

\section{\label{sec:Model} Lattice model of topological chiral magnetic insulators}

The magnetic insulator with broken inversion symmetry
is an ideal playground in which inhomogeneity induced responses can be investigated.
This is because the lack of the spatial inversion symmetry allows Dzyaloshinskii-Moriya
interaction which leads to spatially modulated spin structures.
In this work, we focus on the magnetic insulators which can be obtained spontaneously from TBIs
including inversion symmetry breaking terms.
To demonstrate the main idea concretely,
we start by constructing a model Hamiltonian of TCMI
on the honeycomb lattice.
Up to now, two types of TBIs are proposed on the honeycomb lattice.
One is the anomalous Hall insulator (AHI) with broken time-reversal symmetry
and the other is the time-reversal invariant spin Hall insulator (SHI).
Starting from these TBIs, magnetic ordering can be developed when electron correlation is considered.~\cite{Shitade}

The Hamiltonian describing the TCMI can be written
in the following way,
\begin{align}
\hat{H}=\hat{H}_{1}+\hat{H}_{2}+\hat{H}_{\text{U}},
\end{align}
in which
\begin{align}
\hat{H}_{1}=&t_{1}\sum_{\langle ij\rangle}\sum_{\sigma}[c^{\dag}_{i\sigma}c_{j\sigma}+ h.c.],
\nonumber\\
\hat{H}_{2}=&\sum_{i}\mu_{S,i}c^{\dag}_{i}c_{i}
+\frac{i}{3\sqrt{3}}\sum_{\langle\langle ij\rangle\rangle}\sum_{\sigma=\uparrow,\downarrow}
t_{2}(\sigma)c^{\dag}_{i,\sigma}\hat{z}\cdot\hat{e}_{ij}c_{j,\sigma},
\nonumber
\end{align}
where $\langle ij \rangle$ and $\langle\langle ij \rangle\rangle$
denote the nearest-neighbor (NN) and next-nearest-neighbor (NNN) pairs, respectively.
$t_{1}$ indicates the hopping amplitude between NN sites.
$t_{2}(\sigma)$
is a spin independent constant $t_{2}(\sigma)=t_{2}$ in AHI while
$t_{2}(\sigma)=t_{2}$ $(-t_{2})$ for spin-up (spin-down) electrons in SHI,
where $t_{2}$ is the amplitude of the pure imaginary hopping between NNN sites.
The unit vector $\hat{e}_{ij}$ is defined as
$\hat{e}_{ij}=(\textbf{d}^{1}_{ij}\times \textbf{d}^{2}_{ij})/|\textbf{d}^{1}_{ij}\times \textbf{d}^{2}_{ij}|$
where $\textbf{d}^{1}_{ij}$ and $\textbf{d}^{2}_{ij}$ are the bond unit vectors
along the two bonds which are traversed by the electron when it moves from the site $j$ to $i$.
The noninteracting Hamiltonian $\hat{H}_{1}$
has a semi-metallic ground state at half-filling
with two Dirac points at the momenta $\textbf{K}_{1}$ and $\textbf{K}_{2}(=-\textbf{K}_{1})$.
The staggered chemical potential $\mu_{S,i}$ in $\hat{H}_{2}$, which
breaks the lattice inversion symmetry, satisfies $\mu_{S,i}=\mu_{S}$ ($-\mu_{S}$)
if $i$ belongs to the $A$ ($B$) sublattice.
Considering the potential realization of CMI in
bilayers of the perovskite-type transition metal oxides grown along the [111] direction,
the inclusion of the staggered chemical potential $\mu_{S,i}$ is required.~\cite{111_1,111_2}
Interestingly, it is shown later that the existence of the inversion symmetry breaking term $\mu_{S,i}$ plays
an essential role to stabilize TCMIs.
\begin{figure}[t]
\centering
\includegraphics[width=8.5 cm]{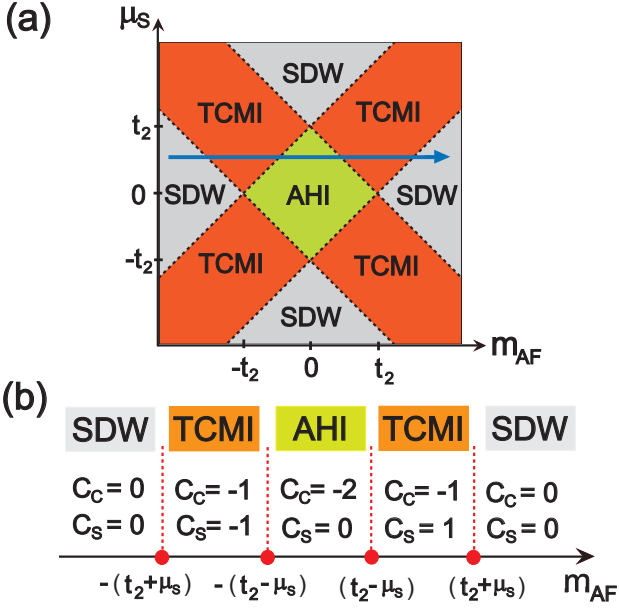}
\caption{(Color online)
The schematic phase diagram of the full lattice mean field
Hamiltonian $\hat{H}_{\text{Full}}$ (Eq.~(\ref{eqn:fullHamiltonian}))
for collinear magnetic phases derived from AHI.
(a) The phase diagram shown in two dimensional space of ($m_{\text{AF}}$, $\mu_{\text{S}}$)
for given $t_{2}>0$.
(b) One dimensional section of the 2D phase diagram, which is obtained by varying $m_{\text{AF}}$
for given $t_{2}>\mu_{\text{S}}>0$ along the long arrow in (a).
In both (a) and (b), a dotted line indicates
the gapless phase boundary between gapped phases.
In (b), each phase is characterized by the charge ($C_{C}$)
and the spin ($C_{S}$) Chern numbers.
Here TCMI (SDW) means a topological chiral magnetic insulator (spin density wave insulator).
} \label{fig:phasediagram_AHI}
\end{figure}
\begin{figure}[t]
\centering
\includegraphics[width=8.5 cm]{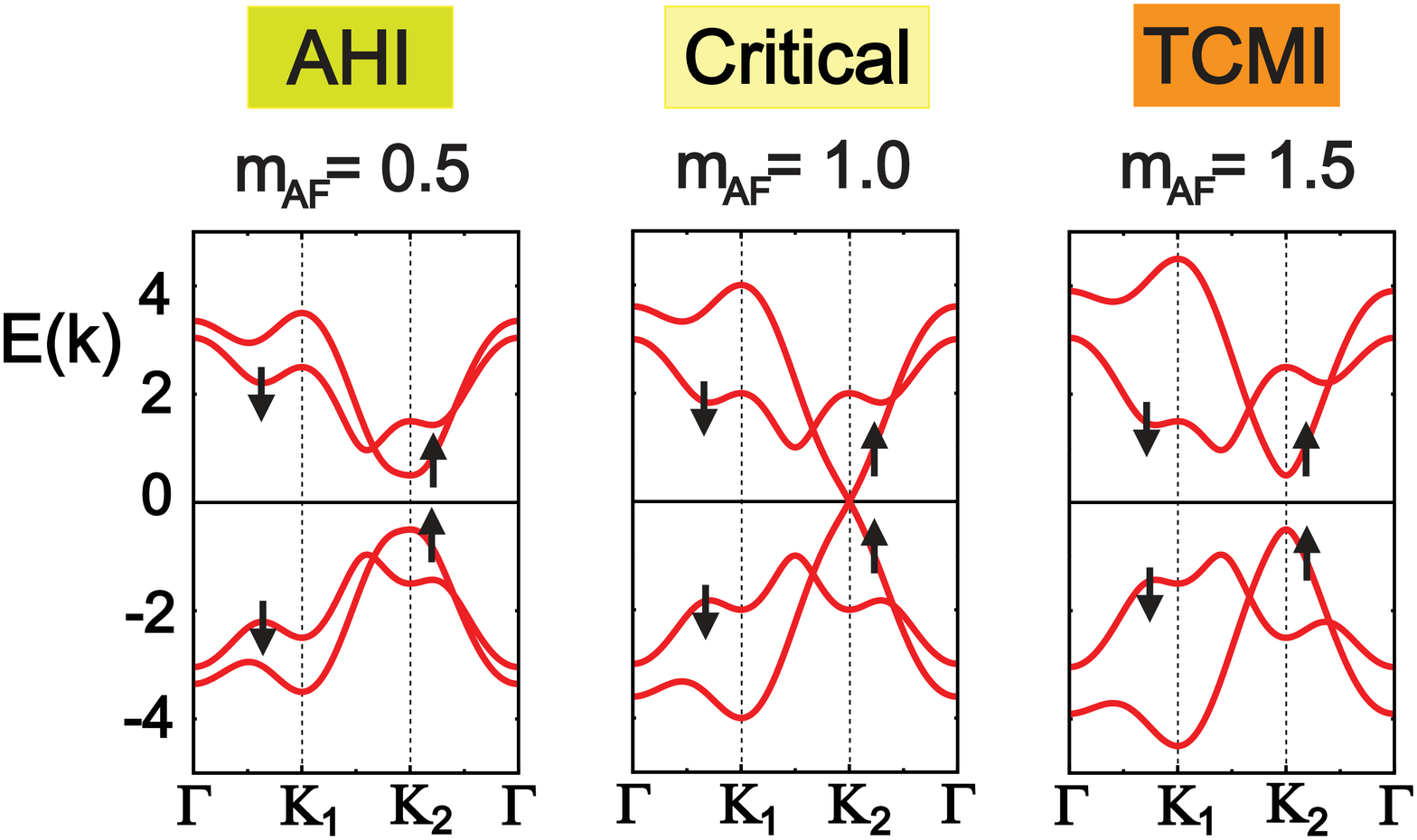}
\caption{(Color online)
Evolution of the band structure as $m_{\text{AF}}$ changes for $t_{2}=2$,
$\mu_{\text{S}}=1$. The AHI ($m_{\text{AF}}=0.5$) turns into
the TCMI ($m_{\text{AF}}=1.5$) via a critical point at $m_{\text{AF}}=t_{2}-\mu_{\text{S}}=1$.
Due to the broken time reversal and inversion symmetries,
a band touching occurs only at $\textbf{k}=\textbf{K}_{2}$ between spin-up bands,
which changes the charge and spin Chern numbers simultaneously.
} \label{fig:bandstructure}
\end{figure}

$\hat{H}_{\text{U}}$ indicates electron-electron interactions generating a magnetic ground state.
We introduce a collinear spin ordering along $\hat{r}$ direction represented by
$\textbf{S}_{A}=m_{A}\hat{r}$
and $\textbf{S}_{B}=m_{B}\hat{r}$ at the two sites $A$ and $B$ in a unit cell.
Assuming the translational invariance of the magnetic ground state,
the mean field approximation for $\hat{H}_{\text{U}}$ leads to the following Hamiltonian,
\begin{align}\label{eqn:H_sdw}
\hat{H}_{\text{MF}}
&=\sum_{\textbf{k}}\sum_{i=x,y,z}\psi^{\dag}(\textbf{k})[m_{\text{F},i}(\tau_{0}\sigma_{i})
+m_{\text{AF},i}(\tau_{z}\sigma_{i})]\psi(\textbf{k}),
\nonumber\\
&=\sum_{\textbf{k}}\psi^{\dag}(\textbf{k})[(\textbf{m}_{\text{F}}\cdot\vec{\sigma})
+\tau_{z}(\textbf{m}_{\text{AF}}\cdot\vec{\sigma})]\psi(\textbf{k}),
\end{align}
where $\psi^{\dag}(\textbf{k})=(c^{\dag}_{A,\uparrow}(\textbf{k}), c^{\dag}_{A,\downarrow}(\textbf{k}),c^{\dag}_{B,\uparrow}(\textbf{k}),c^{\dag}_{B,\downarrow}(\textbf{k}))$.
$\textbf{m}_{\text{F}}=m_{\text{F}}\hat{r}$ and $\textbf{m}_{\text{AF}}=m_{\text{AF}}\hat{r}$
with $m_{\text{F}}=(m_{A}+m_{B})/2$ and $m_{\text{AF}}=(m_{A}-m_{B})/2$.
$\sigma_{i}$ and $\tau_{i}$ ($i=x,y,z$) are Pauli matrices
for spin and sublattice degrees of freedom and $\sigma_{0}$, $\tau_{0}$ are identity matrices.
Here $\sigma_{i}\tau_{j}\equiv\sigma_{i}\otimes\tau_{j}$ indicates
a $4\times 4$ matrix acting on the 4 component vector $\psi(\textbf{k})$.
Also when the direct product of Pauli matrices is involved with an identity matrix
such as $\tau_{0}$ or $\sigma_{0}$,
we omit the identity matrix in the product if the dimensionality of the Hamiltonian is
obvious from the context. For example, $\tau_{0}\sigma_{i}=\sigma_{i}$ in Eq.~(\ref{eqn:H_sdw}).
Assuming $|m_{\text{AF}}|\gg|m_{\text{F}}|$, we first neglect
the ferromagnetic component $m_{\text{F}}$, whose influence on the adiabatic polarization
is considered later.

\begin{figure}[t]
\centering
\includegraphics[width=8.8 cm]{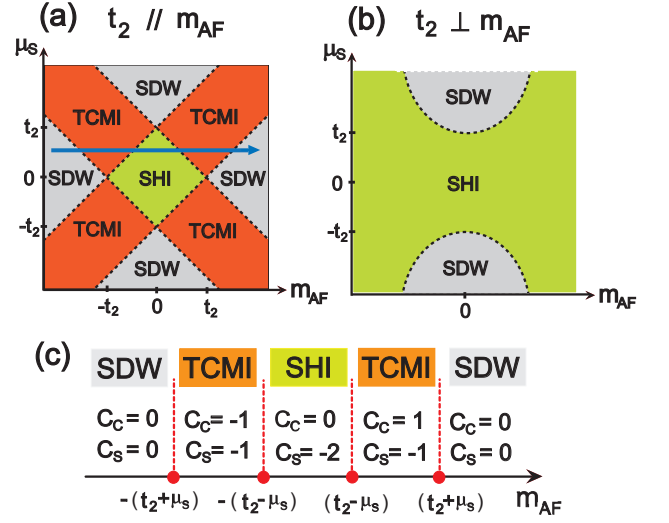}
\caption{(Color online)
The schematic phase diagram of the full lattice mean field
Hamiltonian $\hat{H}_{\text{Full}}$ (Eq.~(\ref{eqn:fullHamiltonian}))
for collinear magnetic phases derived from SHI.
(a) The phase diagram shown in two dimensional space of ($m_{\text{AF}}$, $\mu_{\text{S}}$)
for given $t_{2}>0$ when $\textbf{t}_{2}\parallel \textbf{m}_{\text{AF}}$.
(b) The phase diagram for $\textbf{t}_{2}\perp \textbf{m}_{\text{AF}}$.
(c) One dimensional section of the 2D phase diagram, which is obtained by varying $m_{\text{AF}}$
for given $t_{2}>\mu_{\text{S}}>0$ along the long arrow in (a).
In (a), (b), (c), a dotted line indicates
the gapless phase boundary between two gapped phases.
} \label{fig:phasediagram_SHI}
\end{figure}

The full lattice mean field Hamiltonian can be written as
\begin{align}\label{eqn:fullHamiltonian}
\hat{H}_{\text{Full}}=\hat{H}_{1}+\hat{H}_{2}+\hat{H}_{\text{MF}},
\end{align}
which leads to the phase diagram shown in Fig.~\ref{fig:phasediagram_AHI} and Fig.~\ref{fig:phasediagram_SHI}.
The phase diagram is obtained in the following way.
Since the (topological) phase transition between neighboring insulators
occurs via band touching at $\textbf{k}=\textbf{K}_{1}$ or $\textbf{K}_{2}$,
the phase diagram can be investigated by constructing the effective Hamiltonain
describing the low energy particles near the two nodes $\textbf{k}=\textbf{K}_{1}$ or $\textbf{K}_{2}$.
Explicitly, it is given by
\begin{align}\label{eqn:H_Dirac}
H_{\text{low}}(\textbf{q})=q_{x}\tau_{x}\nu_{z}+q_{y}\tau_{y}+\mu_{\text{S}}\tau_{z}
+t_{2}\tau_{z}\nu_{z}\sigma'_{\alpha}+m_{\text{AF}}\tau_{z}\sigma_{z},
\end{align}
where $\sigma'_{\alpha}=\sigma_{0}$ ($\sigma'_{\alpha}=\sigma_{z}$) for AHI (SHI).
The Pauli matrices $\nu_{x,y,z}$ are introduced to indicate the node degrees of freedom, i.e.,
$\textbf{K}_{1}$ ($\nu_{z}=1$) and $\textbf{K}_{2}$ ($\nu_{z}=-1$).
The momentum $(q_{x},q_{y})$ is measured with respect to the nodes.
When the Hamiltonian can be written as a massive Dirac Hamiltonian
such as $H=q_{x}\tau_{x}+q_{y}\tau_{y}+m\tau_{z}$, the sign reversal
of the mass $m$ changes the Chern number of the occupied band by $\Delta C=\text{sgn}(m)$
via a band touching at $m=0$.
Therefore the location of the phase boundary and
the consequential change of Chern numbers
can be understood by comparing the relative magnitude of $\mu_{\text{S}}$,
$t_{2}$, and $m_{\text{AF}}$ in Eq.~(\ref{eqn:H_Dirac}).
In addition, if the Chern number of one phase is known, the Chern numbers of
all the other phases can be determined by investigating the sign change of
the mass term of the effective Dirac Hamiltonian. For example,
if $m_{\text{AF}}\gg t_{2}$ and $m_{\text{AF}}\gg \mu_{\text{S}}$, the corresponding
gapped phase, i.e., the SDW phase, should have zero Chern number
since it is a topologically trivial. In this way,
the phase diagrams in Fig.~\ref{fig:phasediagram_AHI} and Fig.~\ref{fig:phasediagram_SHI}
can be obtained.

In addition,
to confirm the structure of the phase diagram and topological property
of each phase,
the Chern number
of each phase is numerically computed including
the full band dispersion in the Brillouin zone.
The Chern number $C_{\sigma}$ of the occupied band with spin $\sigma$
is defined as,
\begin{align}\label{eq:chernintegral}
C_{\sigma}\equiv&\frac{1}{2\pi}\int d^{2}k F_{\sigma}(\textbf{k}),
\end{align}
where the momentum space Berry curvature $F_{\sigma}(\textbf{k})$
is defined as
$F_{\sigma}(\textbf{k})\equiv\partial_{k_{x}}A_{y,\sigma}(\textbf{k})-\partial_{k_{y}}A_{x,\sigma}(\textbf{k})$
in which the Berry potential $A_{\mu,\sigma}(\textbf{k})$ is given by
$A_{\mu,\sigma}(\textbf{k})=
-i\langle \Phi_{\sigma}(\textbf{k})|\partial_{k_{\mu}}|\Phi_{\sigma}(\textbf{k})\rangle$.~\cite{TKNN1,TKNN2}
Here $|\Phi_{\sigma}(\textbf{k})\rangle$ indicates the periodic part of the Bloch wave function
corresponding to the occupied state with spin $\sigma$.

Interestingly, various
insulating magnetic phases with distinct topological properties are possible depending
on the relative magnitudes of $m_{\text{AF}}$, $t_{2}$, and $\mu_{\text{S}}$.
Different insulating phases are distinguished based on
the Chern numbers obtained for the collinear magnetic ground states.
Fig.~\ref{fig:phasediagram_AHI} shows the magnetic ground states derived
from AHI. Here we continue to use the term ``AHI"
as long as the magnetic phase possesses the charge Chern number $C_{C}=\pm2$ and the spin Chern number $C_{S}=0$.
The charge ($C_{C}$) and spin ($C_{S}$)
Chern numbers are defined as $C_{C}=C_{\uparrow}+C_{\downarrow}$ and
$C_{S}=C_{\uparrow}-C_{\downarrow}$.
In particular, for $||t_{2}|-|\mu_{S}||<|m_{\text{AF}}|<||t_{2}|+|\mu_{S}||$,
the TCMI, which can be characterized by the odd integer charge and spin Chern
numbers ($|C_{C}|=|C_{S}|=1$), is obtained.
SDW indicates the magnetic insulator with $C_{C}=C_{S}=0$.

The band structures of two topological phases, i.e., AHI and TCMI,
and the band touching at the phase boundary between them are shown in Fig.~\ref{fig:bandstructure}.
It is interesting to note that due to the broken
time reversal and inversion symmetries, the band touching occurs
only between spin-up bands at $\textbf{k}=\textbf{K}_{2}$.
Since the Chern number for the occupied spin-up band changes by 1 ($\Delta C_{\uparrow}=1$),
the charge and spin Chern numbers also change simultaneously leading
to TCMI phase.

In the case of magnetic insulators derived from SHI,
the nature of the ground state depends crucially on the
relative orientation between $\textbf{m}_{\text{AF}}$ and $\textbf{t}_{2}\equiv t_{2}\hat{z}$.
If $\textbf{m}_{\text{AF}} \| \textbf{t}_{2}$, i.e., $\textbf{m}_{\text{AF}}=m_{\text{AF}}\hat{z}$,
the phase diagram is shown in Fig.~\ref{fig:phasediagram_SHI} (a) and (c).
In this case, similar to the case of magnetic insulators derived from AHI,
various topological insulators can be obtained depending on the
relative magnitude of $t_{2}$, $\mu_{\text{S}}$, and $m_{\text{AF}}$.
Here ``SHI" also includes the magnetic phases which have the same Chern
numbers as the usual nonmagnetic SHI phase assuming that the spin $z$ component $S_{z}$
is conserved. However, if $S_{z}$ is not conserved, it is equivalent to a topologically trivial phase.
As the relative angle between $\textbf{m}_{\text{AF}}$ and $\textbf{t}_{2}$ increases,
the phase boundaries between different gapped phases change.
In particular, when $\textbf{m}_{\text{AF}} \bot \textbf{t}_{2}$,
the area of TCMI phase shrinks to zero and the phase diagram contains only two phases, i.e., SHI and SDW.
This is because when $\textbf{m}_{\text{AF}} \bot \textbf{t}_{2}$, in contrast to the case of $\textbf{m}_{\text{AF}} \| \textbf{t}_{2}$,
gap-closing appears at the two nodes simultaneously
along the phase boundary between SHI and SDW. In this case, the Chern number of the system does not change.
The variation of the phase diagram as the relative angle between $\textbf{m}_{\text{AF}}$ and $\textbf{t}_{2}$ changes
implies that
starting from a TCMI phase with given $\textbf{t}_{2}$, $\mu_{\text{S}}$, and $\textbf{m}_{\text{AF}}$,
if we rotate the direction of $\textbf{m}_{\text{AF}}$ relative to $\textbf{t}_{2}$,
topological phase transitions should occur, through which TCMI turns into either SDW or SHI phase.

\begin{figure}[t]
\centering
\includegraphics[width=8.5 cm]{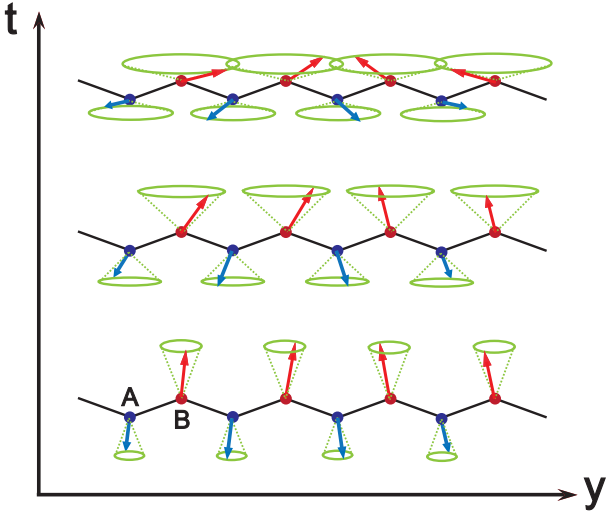}
\caption{(Color online)
An example of the adiabatic variation of the inhomogeneous spin structure.
A spiral spin state with the modulation wave vector $\textbf{Q}=q\hat{y}$
is described. A pair of the spins at the two sublattice sites $A$ and $B$ are
aligned in the opposite directions.
The magnitude of the uniform component of the rotating spins
is varying adiabatically (along the vertical direction in the figure),
which can be controlled in experiment by applying
external magnetic fields.
} \label{fig:spiral}
\end{figure}

From the collinear magnetic states,
the inhomogeneous spin structures can be introduced by
considering smooth deformation of spin ordering directions.
When the inversion symmetry breaking term $\mu_{S}$ is much smaller
than the spin exchange couplings, the spins are aligned almost collinearly.
In this limit, the local inhomogeneous spin structures, such as skyrmion defects and domain walls,
can be developed.
On the other hand, since the Dzyaloshinskii-Moriya interaction is allowed between localized spins
when the inversion symmetry breaking term is finite,
the inhomogeneous spin structure can be developed globally leading
to spiral (or conical) spin orderings.
All these inhomogeneous spin structures can be described by replacing
$m_{\text{AF}}\hat{r}\cdot\vec{\sigma}$ with $m_{\text{AF}}\vec{\sigma}\cdot\hat{N}(\textbf{r},\eta)$
where the three-component unit vector $\hat{N}$ is given by
\begin{align}\label{eqn:spiral}
\hat{N}(\textbf{r},\eta)&\equiv(N_{x},N_{y},N_{z})
\nonumber\\
&=(\sin\theta(\eta)\cos\phi(\textbf{r}),\sin\theta(\eta)\sin\phi(\textbf{r}),\cos\theta(\eta)).
\end{align}
For example,
for $\textbf{m}_{\text{AF}}=m_{\text{AF}}\hat{z}$,
a static configuration with $\theta(\eta)=\theta(|\textbf{r}|)$ satisfying the boundary conditions
$\theta(|\textbf{r}|=0)=\pi$ and $\theta(|\textbf{r}|\rightarrow\infty)=0$,
describes a skyrmion. Adopting $\phi(\textbf{r})=\tan^{-1}(\frac{y}{x})$,
a skyrmion can be specified by the two parameters $(|\textbf{r}|,\phi)$ which
is nothing but the polar coordinates for two dimensional space.
On the other hand, taking $\phi(\textbf{r})=\textbf{Q}\cdot \textbf{r}$ with proper boundary conditions,
$\hat{N}(\textbf{r},\eta)$ describes
a spiral spin ordering with the modulation wave vector $\textbf{Q}$
or a smooth variation of spin directions around a domain wall.~\cite{Maxim}
In these cases, the spatial modulation itself can be described by a single parameter.
For example, taking $\textbf{Q}=Q\hat{y}$, $\phi(\textbf{r})=\phi(y)$
depends only on the $y$-coordinate.
If we adopt the time coordinate $t$ as $\eta$, $\hat{N}(y,t)$
describes the temporal variation of the inhomogeneous spin structures.
Then $N_{z}=\cos\theta(t)$ denotes the uniform spin component
independent of the spatial coordinates and
the adiabatic variation of $\theta(t)$ controls the relative magnitude
of the rotating and uniform spin components.
In Fig.~\ref{fig:spiral}, we describe the adiabatic temporal change
of the spiral ferrimagnetic ordering. Here the magnitude
of the uniform component decreases as a function of $t$.
It is the main purpose of this paper to understand the topological responses of the system due
to the adiabatic change of the $\hat{N}$ which varies in the two dimensional parameter spaces,
either $(|\textbf{r}|,\phi)$ or $(y,t)$.

\section{\label{sec:TCMI_AHI} Topological responses of chiral magnetic insulators derived from AHI}

\subsection{\label{sec:AHI} Effective action for the charge and spin responses}
Let us first consider the charge and spin responses of CMIs which are derived from AHI.
The low energy Hamiltonian, which is obtained by linearizing the energy spectrum of $\hat{H}_{\text{Full}}$
near the two Dirac nodes, can be written as
\begin{align}\label{eqn:effH1}
H_{\text{eff}}=&\int d^{2}r \psi^{\dag}(\textbf{r},t)[(-i\partial_{x})\tau_{x}\nu_{z}+(-i\partial_{y})\tau_{y}
\nonumber\\
&+\mu_{S}\tau_{z}+t_{2}\tau_{z}\nu_{z}
+m_{\text{AF}}\tau_{z}\hat{N}(\textbf{r},t)\cdot\vec{\sigma}]\psi(\textbf{r},t),
\end{align}
where the Fermi velocity of the Dirac particle is scaled to 1 and $\hbar$ is set to 1.
The corresponding action is given by
\begin{align}
S=&\int dtd^{2}r \psi^{\dag}(\textbf{r},t)\gamma^{t}[(-i\gamma^{\mu}\partial_{\mu})
\nonumber\\
&-i\mu_{S}-it_{2}\nu_{z}
-im_{\text{AF}}\hat{N}(\textbf{r},t)\cdot\vec{\sigma}]\psi(\textbf{r},t),
\end{align}
where $\gamma^{\mu}$ ($\mu=t,x,y$) is defined as
$(\gamma^{t},\gamma^{x},\gamma^{y})=(-i\tau_{z}, \tau_{y}\nu_{z}, -\tau_{x})$
and the summation over repeated indices is assumed throughout the paper.

We consider a local unitary transformation $U$ satisfying
$U^{\dag}(\vec{\sigma}\cdot\hat{N})U=\sigma_{z}$,
which rotates each spin to the $+z$ direction.
Defining $\psi \equiv U \psi'$,
the effective action can be written as
\begin{align}
S=&\int dtd^{2}r \psi'^{\dag}(\textbf{r},t)\gamma^{t}[\gamma^{\mu}(-i\partial_{\mu}+A^{C}_{\mu}-\frac{1}{2}\vec{\sigma}\cdot \overrightarrow{B}_{\mu})
\nonumber\\
&-i\mu_{S}-it_{2}\nu_{z}
-im_{\text{AF}}\sigma_{z}]\psi'(\textbf{r},t),
\end{align}
where $\frac{1}{2}\vec{\sigma}\cdot \overrightarrow{B}_{\mu}
\equiv i U^{\dag}(\partial_{\mu}U)$, which shows that the inhomogeneous spin structure
induces SU(2) spin gauge fields in the rotated frame. The electromagnetic
U(1) gauge fields $A^{C}_{\mu}$ are introduced via minimal coupling and
the electron charge $-e$ is set to -1.
To derive the effective action for gauge fields, we split the action
into two pieces $S=S_{1}+S_{2}$ in which
\begin{align}\label{eqn:effectivaction}
S_{1}=&\int \frac{d^{3}k}{(2\pi)^3} \psi'^{\dag}(k)G_{0}^{-1}(k)\psi'(k),
\nonumber\\
S_{2}=&\int \frac{d^{3}k}{(2\pi)^3}\frac{d^{3}q}{(2\pi)^3} \psi'^{\dag}(k+q)
\nonumber\\
&\quad\times\Delta_{\mu}(\partial_{k_{\mu}}G_{0}^{-1})[eA^{C}_{\mu}(q)-\frac{1}{2}\vec{\sigma}\cdot \overrightarrow{B}_{\mu}(q)]\psi'(k),
\end{align}
where
\begin{align}
G_{0}^{-1}(k)=\omega-[k_{x}\tau_{x}\nu_{z}+k_{y}\tau_{y}
+\mu_{S}\tau_{z}+t_{2}\tau_{z}\nu_{z}
+m_{\text{AF}}\tau_{z}\sigma_{z}].
\end{align}
Here the three momenta $k$ is defined as $k=(\omega,k_{x},k_{y})$
and $(\Delta_{t},\Delta_{x},\Delta_{y})=(-1,1,1)$.
The effective action can be obtained after integrating out
the fermion fields and expanding the resulting action
in powers of the gauge fields and their gradients.
Straightforward calculation gives rise to the following expression
for the effective gauge action,~\cite{Yakovenko1,Sengupta}
\begin{align}\label{eqn:Sgauge}
S_{\text{eff}}[A^{C}_{\mu},A^{S}_{\mu}]=&\frac{C_{C}}{4\pi}\int dt d^{2}r \epsilon^{\mu\nu\lambda}
A^{C}_{\mu}\partial_{\nu}A^{C}_{\lambda}
\nonumber\\
&+\frac{C_{S}}{4\pi}\int dt d^{2}r \epsilon^{\mu\nu\lambda}
A^{S}_{\mu}\partial_{\nu}A^{C}_{\lambda}
\nonumber\\
&+\frac{C_{C}}{16\pi}\int dt d^{2}r \epsilon^{\mu\nu\lambda}
A^{S}_{\mu}\partial_{\nu}A^{S}_{\lambda},
\end{align}
where
\begin{align}\label{eqn:ChernNumbers}
C_{C}=&\frac{\epsilon^{\mu\nu\lambda}}{24\pi^2}\int d^{3}k
\text{Tr}[G_{0}\frac{\partial G_{0}^{-1}}{\partial k_{\mu}}G_{0}\frac{\partial G_{0}^{-1}}{\partial k_{\nu}}G_{0}
\frac{\partial G_{0}^{-1}}{\partial k_{\lambda}}],
\nonumber\\
C_{S}=&\frac{\epsilon^{\mu\nu\lambda}}{24\pi^2}\int d^{3}k
\text{Tr}[\sigma_{z}G_{0}\frac{\partial G_{0}^{-1}}{\partial k_{\mu}}G_{0}\frac{\partial G_{0}^{-1}}{\partial k_{\nu}}G_{0}
\frac{\partial G_{0}^{-1}}{\partial k_{\lambda}}],
\end{align}
which are nothing but the charge ($C_{C}$) and spin ($C_{S}$) first Chern numbers
of the collinear magnetic ground state described by $G_{0}$.
The U(1) spin gauge field $A_{\mu}^{S}$ is defined as $A_{\mu}^{S}\equiv B_{\mu,z}$.
It is straightforward to show that the fictitious spin gauge flux $F^{S}_{\mu\nu}\equiv\partial_{\mu}A^{S}_{\nu}-\partial_{\nu}A^{S}_{\mu}$
satisfies
\begin{align}\label{eqn:spinflux}
F^{S}_{\mu\nu}=\partial_{\mu}A^{S}_{\nu}-\partial_{\nu}A^{S}_{\mu}=
\hat{N}\cdot[(\partial_{\mu}\hat{N})\times(\partial_{\nu}\hat{N})].
\end{align}

The effective gauge action in Eq.~(\ref{eqn:Sgauge}) gives rise to the
following expressions for the charge current,
\begin{align}\label{eqn:chargecurrent}
j^{C}_{\mu}(r,t)&=\frac{\partial S_{\text{eff}}[A^{C}_{\mu},A^{S}_{\mu}]}{\partial A^{C}_{\mu}(r,t)},
\nonumber\\
&=\frac{C_{C}}{2\pi} \epsilon^{\mu\nu\lambda}\partial_{\nu}A^{C}_{\lambda}
+\frac{C_{S}}{4\pi}\epsilon^{\mu\nu\lambda}\partial_{\nu}A^{S}_{\lambda},
\end{align}
and the spin current,
\begin{align}\label{eqn:spincurrent}
j^{S}_{\mu}(r,t)&=\frac{\partial S_{\text{eff}}[A^{C}_{\mu},A^{S}_{\mu}]}{\partial A^{S}_{\mu}(r,t)},
\nonumber\\
&=\frac{C_{S}}{4\pi}\epsilon^{\mu\nu\lambda}\partial_{\nu}A^{C}_{\lambda}+
\frac{C_{C}}{8\pi}\epsilon^{\mu\nu\lambda}\partial_{\nu}A^{S}_{\lambda}.
\end{align}

In the absence of the spin gauge flux $F^{S}_{\mu\nu}$, for example, for a homogeneous
collinear magnetic ground state, the above
charge and spin currents describe the usual quantized charge and spin Hall effects
driven by the external electromagnetic fields.
The magnitude of the quantized charge (spin) current
is determined by the charge (spin) Chern number.
On the other hand,
if the spin gauge flux can be generated by space-time dependent
variation of the order parameter $\hat{N}$ through the relation in Eq.~(\ref{eqn:spinflux}),
the adiabatic charge and spin Hall currents flow in the absence of the electromagnetic fields.
In this case, interestingly, the magnitude of the quantized charge (spin) current
is determined by the spin (charge) Chern number in contrast to the case
of the usual charge and spin Hall currents.

\subsection{\label{sec:DW_AHI} Charge and spin pumping through a domain wall with spiral spins}

\begin{figure}[t]
\centering
\includegraphics[width=8.5 cm]{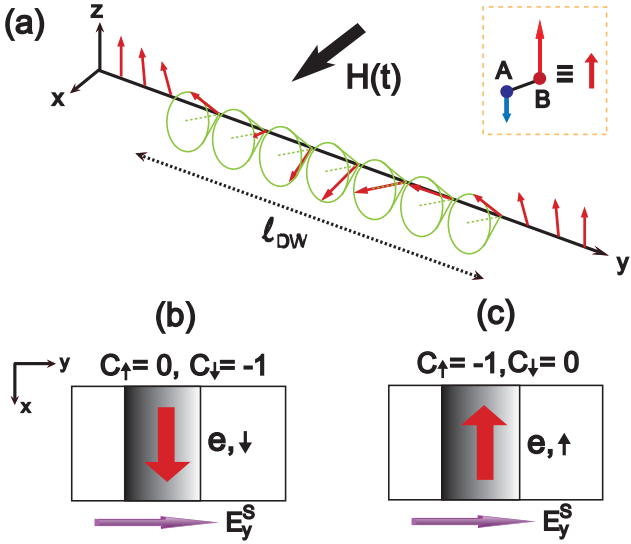}
\caption{(Color online)
Charge and spin currents driven by the spin gauge flux
which is produced by the slow variation of the uniform ($x$) component of the spins rotating in $yz$ plane within
a domain wall.
(a) The rotating spins near the in-phase domain wall.
Here the red rotating arrow describes the net ferromagnetic moment
of a pair of the antiparallel spins within a unit cell.
The external time dependent magnetic field $\textbf{H}(t)=H(t) \hat{x}$
couples to the net ferromagnetic moment and controls the magnitude of the $x$-component
of the rotating spins.
In (b) and (c),
the central square with graded colors in each panel,
indicates the spiral domain wall producing the spin gauge flux on which the direction
of the induced current is denoted with a red arrow.
For a given spin gauge flux
$E^{S}_{y}$, the direction of the adiabatic charge (spin) current is determined by
the corresponding spin (charge) Chern number.
(b) A TCMI with $(C_{S},C_{C})=(1,-1)$.
(c) A TCMI with $(C_{S},C_{C})=(-1,-1)$.
}\label{fig:spincurrent}
\end{figure}

To demonstrate the idea of the spin gauge flux induced charge and spin responses,
we consider the charge and spin polarizations induced by
the adiabatic modulation of the spin direction in a domain wall of a CMI,
which is described in Fig.~\ref{fig:spincurrent}(a).
An in-phase domain wall can be represented by $\hat{N}(\textbf{r},\eta)$ in Eq.~(\ref{eqn:spiral})
with $\phi(\textbf{r})=\phi(y)=\frac{\pi}{2}[-1-2\tanh(y/\ell_{\text{DW}})]$.
For convenience, we rearrange the components of $\hat{N}$ and consider $\hat{N}=\hat{N}_{X}(\textbf{r},t)$
given by
\begin{align}\label{eqn:Xspiral}
\hat{N}_{X}(\textbf{r},t)
&=(\cos\theta(t),\sin\theta(t)\cos\phi(y),\sin\theta(t)\sin\phi(y)),
\end{align}
where the spin $x$-component is uniform.
Traversing the domain wall along the $y$ direction, the spins rotate by $2\pi$ within the $yz$-plane.
External time dependent magnetic field $\textbf{H}(t)=H(t)\hat{x}$ coupled to the spins
within the domain wall controls the magnitude of the uniform ($x$)
components of the rotating spins.
We assume that the length scale of the domain wall $\ell_{\text{DW}}$ satisfies
$\hbar \upsilon/E_{g} \ll \ell_{\text{DW}}\ll L_{y}$. Here
$L_{y}$ represents the system size along the $y$-direction, $E_{g}$ is the bulk energy gap,
and $\upsilon$ is the typical velocity scale contained in the
bulk band structure near the chemical potential.
Therefore the spin modulation near the domain wall is smooth enough
but, overall, the spins are aligned
collinearly throughout the entire system in which the spin $z$ component $S_{z}$ can be treated as a conserved quantity.

Let us first consider the charge polarization induced by the spin gauge flux.
From Eq.~(\ref{eqn:spinflux}) and Eq.~(\ref{eqn:chargecurrent}),
the adiabatic charge polarization current is given by
\begin{align}\label{eqn:chargeHall}
j^{C}_{\mu}&=\frac{C_{S}}{8\pi}\epsilon^{\mu\nu\lambda} \hat{N}\cdot[(\partial_{\nu}\hat{N})\times(\partial_{\lambda}\hat{N})],
\nonumber\\
&=\frac{C_{S}}{4\pi}\delta_{\mu,x} \hat{N}\cdot[(\partial_{y}\hat{N})\times(\partial_{t}\hat{N})].
\end{align}
During the adiabatic evolution of $\theta(t)$ for $t\in[0,T]$,
the spin electric field $E^{S}_{y}\equiv\partial_{y}A^{S}_{t}-\partial_{t}A^{S}_{y}$
=$ \hat{N}\cdot[(\partial_{y}\hat{N})\times(\partial_{t}\hat{N})]$ produced by the spiral spins at the domain wall
induces a charge current $j^{C}_{x}$, which leads to the average charge polarization
$\langle P_{x}\rangle=\frac{1}{L_{y}}\int_{0}^{L_{y}}dy\int_{0}^{T}dt j^{C}_{x}$.
In particular, if the unit vector $\hat{N}(y,t)$ fully covers the surface of the unit sphere
in the $(y,t)$ space during the adiabatic evolution,
a quantized charge pumping is possible
with the net charge transferred to the $x$ direction given by
\begin{align}\label{eqn:chargequantum}
Q_{\text{pumped}}=&\int_{0}^{T}dt\int dy j^{C}_{x}
\nonumber\\
=&\frac{C_{S}}{4\pi}\int_{0}^{\pi}d\theta\int_{\pi/2}^{-3\pi/2} d\phi
\hat{N}\cdot[(\partial_{\phi}\hat{N})\times(\partial_{\theta}\hat{N})],
\nonumber\\
\equiv&C_{S}n_{\text{skyrmion}},
\end{align}
where $n_{\text{skyrmion}}$ is the skyrmion number of the vector field $\hat{N}(y,t)$.
Therefore the unit pumped charge is determined by the product of two topological
invariants defined in momentum space ($C_{S}$) and position/time space ($n_{\text{skyrmion}}$).

Moreover, since an electron carries
both charge and spin quantum numbers,
a charge current accompanies a spin current simultaneously.
The spin current induced by the adiabatic spin electric field can be
understood in the following way.
Since $S_{z}$ is a good quantum number,
a system with $(C_{\uparrow}, C_{\downarrow})=(n_{\uparrow},n_{\downarrow})$
can be considered as a superposition of two subsystems
with $(C_{\uparrow}, C_{\downarrow})=(n_{\uparrow},0)$
and $(C_{\uparrow}, C_{\downarrow})=(0,n_{\downarrow})$, respectively.
For given spin electric field $E^{S}_{y}$, the charge current induced in each subsystem should be
$j^{C}_{x}=e\frac{(n_{\uparrow})}{4\pi}E^{S}_{y}$ and
$e\frac{(-n_{\downarrow})}{4\pi}E^{S}_{y}$, respectively, considering the spin Chern number in each case.
However, since the spin polarization directions are opposite in two subsystems,
the total induced spin current is given by
$j^{S}_{x}=\frac{\hbar}{2}\frac{(n_{\uparrow})}{4\pi}E^{S}_{y}+(-\frac{\hbar}{2})\frac{(-n_{\downarrow})}{4\pi}E^{S}_{y}$
=$\frac{\hbar}{2}\frac{(n_{\uparrow}+n_{\downarrow})}{4\pi}E^{S}_{y}$.
Therefore the amount of spin transfer is determined by the charge Chern number $C_{C}$.
The adiabatic spin current induced by the spin gauge flux can be written as
\begin{align}\label{eqn:spinHall}
j^{S}_{\mu}
=\frac{C_{C}}{16\pi}\epsilon^{\mu\nu\lambda}\hat{N}\cdot[(\partial_{\nu}\hat{N})\times(\partial_{\lambda}\hat{N})].
\end{align}
The identical expression of the spin current can also be obtained from Eq.~(\ref{eqn:spinflux}) and Eq.~(\ref{eqn:spincurrent}).
For an adiabatic cycle in which $\hat{N}(y,t)$ covers the full spherical solid angle,
the net transferred spin along the $x$-direction is given by
\begin{align}\label{eqn:spinquantum}
S_{\text{pumped}}=&\int_{0}^{T}dt\int dy j^{S}_{x}
\nonumber\\
=&\frac{C_{C}}{8\pi}\int_{0}^{\pi}d\theta\int_{\pi/2}^{-3\pi/2} d\phi
\hat{N}\cdot[(\partial_{\phi}\hat{N})\times(\partial_{\theta}\hat{N})],
\nonumber\\
=&\frac{\hbar}{2}C_{C}n_{\text{skyrmion}}.
\end{align}
Therefore the quantum of the pumped spin is determined by
the product of the charge Chern number $C_{C}$ and the skyrmion number $n_{\text{skyrmion}}$.
Interestingly, the charge and spin responses
under the spin gauge fields are exactly dual to the corresponding
responses under the electromagnetic fields.
Given the spin gauge fields, the charge (spin) response
is determined by the spin (charge) Chern number.~\cite{Yakovenko1}

In Fig.~\ref{fig:spincurrent} (b) and (c), we show the adiabatic charge and spin
currents induced by
the adiabatic spin rotation at the spiral domain wall
for the two TCMIs with $C_{S}=1$ and -1 in Fig.~\ref{fig:phasediagram_AHI} (b).
Interestingly, while the charge currents flow in the opposite directions
because their $C_{S}$ have opposite signs,
the directions of the spin currents are the same in the two cases,
consistent with the fact that $C_{C}=-1$ in both cases.
Adiabatic charge and spin currents for AHI can be obtained
by adding the corresponding currents for the two TCMIs, i.e.
by the superposition of Fig.~\ref{fig:spincurrent} (b) and (c).
Since the charge currents flow in the opposite directions,
the net charge polarization is zero. However,
the magnitude of the spin polarization is twice compared to
the case of each TCMI.

\subsection{\label{sec:Skyrmion_AHI} Charge and spin quantum numbers of a skyrmion}

It is worth noting that the last term of the effective gauge action in Eq.~(\ref{eqn:Sgauge}) is nothing
but the Hopf term, which determines
the spin and statistics of skyrmion solitons.~\cite{Wilczek}
Therefore the topological invariants, which quantify the pumped spin and charge,
also characterize the quantum numbers of skyrmion topological textures.~\cite{Teo}
The time components of the adiabatic currents in Eq.~(\ref{eqn:chargeHall}) and Eq.~(\ref{eqn:spinHall})
show the charge and spin densities associated with a skyrmion, which lead
to the following expressions of the charge and spin quantum numbers of a skyrmion,
\begin{align}
Q_{\text{skyrmion}}&=\frac{C_{S}}{4\pi}\int dxdy \hat{N}\cdot[(\partial_{x}\hat{N})\times(\partial_{y}\hat{N})],
\nonumber\\
S_{\text{skyrmion}}&=\frac{C_{C}}{8\pi}\int dxdy \hat{N}\cdot[(\partial_{x}\hat{N})\times(\partial_{y}\hat{N})].
\end{align}
Simple integration of the above equations shows that
a skyrmion with the unit Pontryagin index on the TCMI
is a spin-1/2 fermion with the charge $\pm e$.
On the other hand, a skyrmion with the unit Pontryagin index on the AHI
is a spin-1 boson with no electric charge.
Therefore the skyrmion defects in TCMIs and AHIs carry nontrivial quantum numbers.

\section{\label{sec:TCMI_SHI} Topological responses of chiral magnetic insulators derived from SHI}

\subsection{\label{sec:SHI} Effective action for the charge and spin responses}

The low energy Hamiltonian of a CMI which is derived from the SHI is given by
\begin{align}\label{eqn:effH2}
H_{\text{eff}}=&\int d^{2}r \psi^{\dag}(\textbf{r},t)[(-i\partial_{x})\tau_{x}\nu_{z}+(-i\partial_{y})\tau_{y}
\nonumber\\
&+\mu_{S}\tau_{z}+t_{2}\tau_{z}\nu_{z}\sigma_{z}
+m_{\text{AF}}\tau_{z}\hat{N}(\textbf{r},t)\cdot\vec{\sigma}]\psi(\textbf{r},t).
\end{align}
In contrast to the case of CMI obtained from the AHI,
the local SU(2) spin rotation of the staggered spin order parameter $\hat{N}(\textbf{r},t)$
cannot make the Hamiltonian to be diagonal in spin space
because of the simultaneous rotation of the spin dependent $t_{2}$ term representing the strength of
the spin-orbit coupling.
To obtain the effective gauge action through the gradient expansion, it is necessary
to transfer the position/time dependance of the Hamiltonian to the gauge fields
via local unitary transformations, which can be achieved
only in the limit of $|m_{\text{AF}}|\ll |t_{2}|$
or $|m_{\text{AF}}|\gg |t_{2}|$ corresponding to
SHI or SDW phases in Fig.~\ref{fig:phasediagram_SHI} (c), respectively.
Therefore we focus on the adiabatic responses of SHI and SDW phases
in the following discussion.
Moreover, since the spin is not conserved in general due to the spin-orbit coupling,
we only consider the charge polarization currents.

We first divide the Hamiltonian into two parts in such a way as
$H_{\text{eff}}=H_{\text{eff,+}}+H_{\text{eff,-}}$ in which
$H_{\text{eff,+}}$ ($H_{\text{eff,-}}$) corresponds to $\nu_{z}=1$ ($\nu_{z}=-1$)
describing the low energy particles
near the node at $\textbf{k}=\textbf{K}_{1}$ ($\textbf{k}=-\textbf{K}_{1}$).
Explicitly, $H_{\text{eff},\pm}$ are given by
\begin{align}\label{eqn:Hplusminus}
H_{\text{eff},\pm}=&\int d^{2}r \psi^{\dag}_{\pm}(\textbf{r},t)[\pm(-i\partial_{x})\tau_{x}+(-i\partial_{y})\tau_{y}
\nonumber\\
&+\mu_{S}\tau_{z}+\tau_{z}\{\pm t_{2}\sigma_{z}
+m_{\text{AF}}\hat{N}(\textbf{r},t)\cdot\vec{\sigma}\}]\psi_{\pm}(\textbf{r},t).
\end{align}
To make the Hamiltonian to be diagonal in spin space at each node separately,
we define the collective spin degrees of freedom $\hat{N}_{\pm}$ which combine $\hat{N}\cdot\vec{\sigma}$ and $t_{2}\sigma_{z}$
in the following way,
\begin{align}
&\pm t_{2}\sigma_{z}+m_{\text{AF}}\hat{N}(\textbf{r},t)\cdot\vec{\sigma},
\nonumber\\
&=m_{\text{AF}}N_{x}\sigma_{x}+m_{\text{AF}}N_{y}\sigma_{y}+(m_{\text{AF}}N_{z}\pm t_{2})\sigma_{z},
\nonumber\\
&\equiv \sqrt{m^{2}_{\text{AF}}+t^{2}_{2}\pm 2m_{\text{AF}}t_{2}N_{z}} \hat{N}_{\pm}(\textbf{r},t)\cdot \vec{\sigma},
\end{align}
where the new unit vector $\hat{N}_{\pm}$ satisfies
\begin{align}\label{eqn:skyrmiondensity_SHI}
&\hat{N}_{\pm}\cdot[(\partial_{\mu}\hat{N}_{\pm})\times(\partial_{\nu}\hat{N}_{\pm})]
\nonumber\\
&=\frac{m^{2}_{\text{AF}}\{(m_{\text{AF}}\hat{N}\pm t_{2}\hat{z})\cdot[(\partial_{\mu}\hat{N})\times(\partial_{\nu}\hat{N})]\}}
{[m^{2}_{\text{AF}}+t^{2}_{2}\pm 2m_{\text{AF}}t_{2}N_{z}]^{3/2}}.
\end{align}


Now we consider the local unitary rotations $U_{\pm}$ of the unit vectors $\hat{N}_{\pm}$ satisfying
$U^{\dag}_{\pm}(\vec{\sigma}\cdot \hat{N}_{\pm})U_{\pm}=\sigma_{z}$.
Defining $\psi_{\pm}\equiv U_{\pm}\psi'_{\pm}$, the effective action can
be summarized as that in Eq.~(\ref{eqn:effectivaction})
but with $G_{0}^{-1}$ replaced by
\begin{align}\label{eqn:G0_SHI}
G_{0}^{-1}(k)=&w-(\pm)k_{x}\tau_{x}-k_{y}\tau_{y}-\mu_{S}\tau_{z}
\nonumber\\
&-\sqrt{m^{2}_{\text{AF}}+t^{2}_{2}\pm 2m_{\text{AF}}t_{2}N_{z}} \tau_{z}\sigma_{z},
\end{align}
and $\frac{1}{2}\vec{\sigma}\cdot \overrightarrow{B}_{\mu}
\equiv i U_{+}^{\dag}(\partial_{\mu}U_{+})$ or $i U_{-}^{\dag}(\partial_{\mu}U_{-})$.
In Eq.~(\ref{eqn:G0_SHI}), $G_{0}$ depends on the position/time coordinates due to
the presence of $N_{z}$ term. Therefore when the variation of $G_{0}$ in the position/time
spaces becomes important,
the gradient of $G_{0}$ can also make a nontrivial contribution to the charge polarization
as the gradient of the charge/spin gauge fields does.
Therefore the effective gauge action approach in the previous section is valid only when
$|m_{\text{AF}}|\ll |t_{2}|$ or $|m_{\text{AF}}|\gg |t_{2}|$ is satisfied,
in which $N_{z}$ has a negligible contribution to $G_{0}$.
However, regarding the charge polarization currents shown in Eq.~(\ref{eqn:Sgauge})
and Eq.~(\ref{eqn:ChernNumbers}), $G_{0}$ appears only
in the definition of the Chern number, which is topologically invariant.
Therefore as long as $N_{z}$ does not induce the change of the Chern numbers via band gap-closings,
$N_{z}$ dependence of $G_{0}$ does not affect the charge polarization current obtained by considering the gradients of gauge fields.

Through the gradient expansion of gauge fields following the same procedure as in Sec.~\ref{sec:AHI},
the adiabatic charge current can be obtained as
\begin{align}\label{eqn:currents_SHI}
j^{C}_{\pm,\mu}(\textbf{r},t)&=\frac{C_{S,\pm}}{8\pi}\varepsilon_{\mu\nu\lambda}
\hat{N}_{\pm}\cdot[(\partial_{\nu}\hat{N}_{\pm})\times(\partial_{\lambda}\hat{N}_{\pm})],
\end{align}
where $C_{S,\pm}=\mp 1$ for both $|m_{\text{AF}}|> ||t_{2}|+|\mu_{S}||$ and $|m_{\text{AF}}|< ||t_{2}|-|\mu_{S}||$ cases
corresponding to SDW and SHI phases, respectively.
The total charge polarization of the system can be obtained by adding the adiabatic currents
from the two nodes,
\begin{align}\label{eqn:netcurrents_SHI}
j^{C}_{\mu}(\textbf{r},t)=j^{C}_{+,\mu}(\textbf{r},t)+j^{C}_{-,\mu}(\textbf{r},t).
\end{align}
The adiabatic charge polarization (and also the skyrmion charge) for arbitrary $|m_{\text{AF}}|$, especially for
the TCMI phase corresponding to $||t_{2}|-|\mu_{S}||<|m_{\text{AF}}|<||t_{2}|+|\mu_{S}||$,
is investigated in Sec.~\ref{sec:Semiclassical} by considering
the systematic semiclassical expansion for adiabatic topological responses of inhomogeneous crystals.


\subsection{\label{sec:DW_SHI} Charge polarization at a domain wall with spiral spins}
\begin{figure}[t]
\centering
\includegraphics[width=8.5 cm]{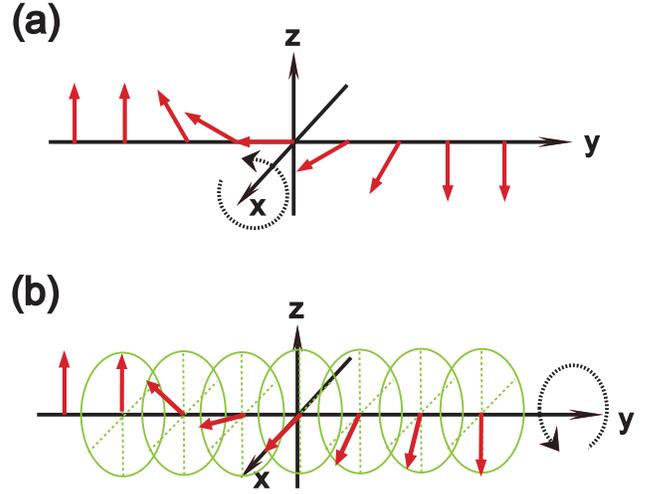}
\caption{(Color online)
Spin structures at anti-phase domain walls for $\textbf{m}_{\text{AF}}=m_{\text{AF}}\hat{z}$.
Here the arrows indicate the staggered spins of the antiferromagnet (or ferrimagnet).
Spins rotate with respect to an axis, $\hat{x}$ or $\hat{y}$, which is normal to the collinear spin direction $\hat{z}$.
(a) The Neel domain wall in which spins rotate around $\hat{x}$ axis, normal
to the spin modulation direction $\hat{y}$.
The spin structure can be described by using $\hat{N}^{\text{DW}}_{X}$ in Eq.~(\ref{eqn:XYZspiral}).
(b) The Bloch domain wall in which spins rotate around $\hat{y}$ axis, parallel
to the spin modulation direction.
The spin structure can be described by using $\hat{N}^{\text{DW}}_{Y}$ in Eq.~(\ref{eqn:XYZspiral}).
} \label{fig:DomainWall}
\end{figure}
We consider the charge polarization or pumping through a domain wall
with spiral spins.
Because of the spin anisotropy due to the spin-orbit interaction $\textbf{t}_{2}=t_{2}\hat{z}$,
the relative orientations of $\textbf{m}_{\text{AF}}$, $\textbf{t}_{2}$, and the rotation axis
of domain wall spins are crucial ingredients determining the charge polarization.
To describe the domain wall spins, we consider the following representations
of inhomogeneous spins,
\begin{align}\label{eqn:XYZspiral}
\hat{N}^{\text{DW}}_{X}(\textbf{r},t)
&=(\cos\theta(t),\sin\theta(t)\cos\phi(y),\sin\theta(t)\sin\phi(y)),
\nonumber\\\
\hat{N}^{\text{DW}}_{Y}(\textbf{r},t)
&=(\sin\theta(t)\sin\phi(y),\cos\theta(t),\sin\theta(t)\cos\phi(y)),
\nonumber\\
\hat{N}^{\text{DW}}_{Z}(\textbf{r},t)
&=(\sin\theta(t)\cos\phi(y),\sin\theta(t)\sin\phi(y),\cos\theta(t)).
\end{align}
In $\hat{N}^{\text{DW}}_{\alpha}$, the spin $\alpha$-component is $\cos\theta(t)$ which is uniform independent of $y$.
When we describe a domain wall with $\hat{N}^{\text{DW}}_{\alpha}$, $\alpha$ indicates the rotation axis of the domain wall spins.
The deviation of $\theta$ from $\frac{\pi}{2}$ describes the adiabatic change
of the relative magnitude of the uniform and rotating components of the domain wall spins.

In Fig.~\ref{fig:DomainWall}, anti-phase domain walls with rotating spins are described for $\textbf{m}_{\text{AF}}=m_{\text{AF}}\hat{z}$.
Here $\hat{N}^{\text{DW}}_{X}$ with the boundary conditions of $\phi(y\rightarrow -\infty)=\frac{\pi}{2}$
and $\phi(y\rightarrow +\infty)=\frac{3\pi}{2}$, describes the Neel domain wall in which the spins rotate around $\hat{x}$-axis perpendicular
to the spin modulation direction $\hat{y}$ as shown in Fig.~\ref{fig:DomainWall} (a).
On the other hand, $\hat{N}^{\text{DW}}_{Y}$ with the boundary conditions of $\phi(y\rightarrow -\infty)=0$
and $\phi(y\rightarrow +\infty)=\pi$, describes the Bloch domain wall in which the spins rotate around
the spin modulation direction $\hat{y}$ as shown in Fig.~\ref{fig:DomainWall} (b).
For these two configurations we compute the partial polarization $\frac{dP_{x}}{d\theta}$
that is defined as
\begin{align}\label{eqn:partialpolarization}
\frac{dP_{x}}{d\theta}=\int_{\phi(y\rightarrow -\infty)}^{\phi(y\rightarrow +\infty)}d\phi j^{C}_{x}(\theta,\phi).
\end{align}
Using Eq.~(\ref{eqn:skyrmiondensity_SHI}), Eq.~(\ref{eqn:currents_SHI}),
and Eq.~(\ref{eqn:netcurrents_SHI}), it can be easily shown that
$\frac{dP_{x}}{d\theta}$ for both $\hat{N}^{\text{DW}}_{X}$
and $\hat{N}^{\text{DW}}_{Y}$ are zero for any $\theta$. Therefore no charge polarization occurs
at the domain walls with rotating spins when $\textbf{m}_{\text{AF}} \parallel \textbf{t}_{2}$.

\begin{figure}[t]
\centering
\includegraphics[width=8.5 cm]{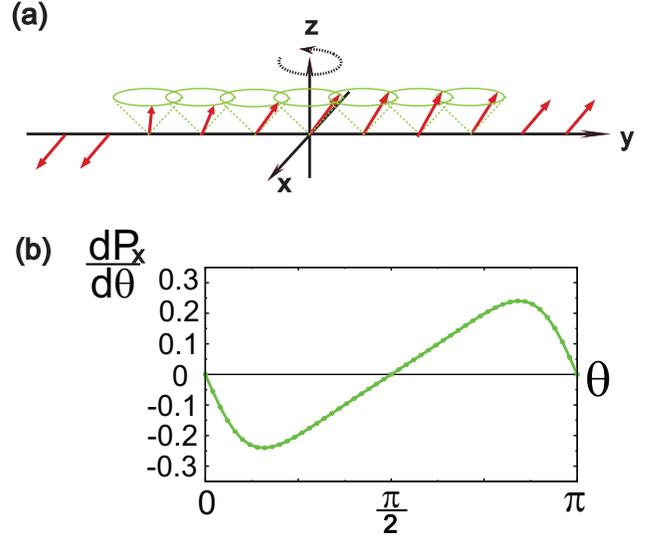}
\caption{(Color online)
Charge polarization at the Neel domain wall described
by $\hat{N}^{\text{DW}}_{Z}=(\sin\theta(t)\cos\phi(y),\sin\theta(t)\sin\phi(y),\cos\theta(t))$
for $\textbf{m}_{\text{AF}}\parallel \hat{x}$. Here the boundary condition
is given by $\phi(y\rightarrow -\infty)=0$ and $\phi(y\rightarrow \infty)=\pi$.
(a) Spin configuration near the Neel domain wall.
(b) $\frac{dP_{x}}{d\theta}$ as a function of $\theta$.
}\label{fig:DW_polarization}
\end{figure}

Similarly, the partial polarization $\frac{dP_{x}}{d\theta}$ induced
by the rotating spins at the domain walls can be considered for $\textbf{m}_{\text{AF}} \perp \textbf{t}_{2}$.
If $\textbf{m}_{\text{AF}}\parallel\hat{x}$, $\hat{N}^{\text{DW}}_{Z}$ ($\hat{N}^{\text{DW}}_{Y}$) describes
the Neel (Bloch) domain wall with appropriate boundary conditions.
It can be shown that $\frac{dP_{x}}{d\theta}\neq0$ in general for the Neel domain wall ($\hat{N}^{\text{DW}}_{Z}$).
On the other hand, in the case of the Bloch domain wall described by $\hat{N}^{\text{DW}}_{Y}$,
$\frac{dP_{x}}{d\theta}$ has opposite signs for two inequivalent domain
wall configurations in which $\phi(y)\in[-\frac{\pi}{2},\frac{\pi}{2}]$
or $\phi(y)\in[\frac{\pi}{2},\frac{3\pi}{2}]$, respectively.
Therefore the average polarization for this case is expected to be zero
when two anti-phase domain wall configurations are equally distributed in material.
If $\textbf{m}_{\text{AF}}\parallel\hat{y}$,
both $\hat{N}^{\text{DW}}_{X}$ and $\hat{N}^{\text{DW}}_{Z}$ describe the Neel domain walls.
For the Neel domain wall described by $\hat{N}^{\text{DW}}_{Z}$,$\frac{dP_{x}}{d\theta}\neq0$ in general.
However, in the case of the domain wall described by $N^{\text{DW}}_{X}$,
$\frac{dP_{x}}{d\theta}$ has opposite signs for two inequivalent domain
wall configurations in which $\phi(y)\in[0,\pi]$ or $\phi(y)\in[\pi,2\pi]$, respectively.
The average polarization for this case is also expected to be zero.

Therefore the partial charge polarization $\frac{dP_{x}}{d\theta}$
can be nonzero only for $\textbf{m}_{\text{AF}} \perp \textbf{t}_{2}$ at the Neel domain walls described by $\hat{N}^{\text{DW}}_{Z}$.
$\frac{dP_{x}}{d\theta}$ from the Neel domain wall is shown in Fig.~\ref{fig:DW_polarization}.
Interestingly, if we assume that the spin rotation occurs over the whole lattice system,
$\hat{N}^{\text{DW}}_{Z}$ ($\hat{N}^{\text{DW}}_{Y}$) describes the transverse (longitudinal) conical magnetic ground state.~\cite{Tokura}
Therefore the charge polarization occurs only for the transverse conical magnetic phases
in the direction perpendicular to the spatial modulation,
consistent with the general theory for the charge polarization in spiral magnets.~\cite{Maxim,Katsura}

Now let us discuss about the quantized charge pumping (also the charge quantum number of a skyrmion) considering
the variations of $\phi\in[0,2\pi]$ and $\theta(t)\in [0,\pi]$.
At first, for $\hat{N}^{\text{DW}}_{Z}$, the local adiabatic charge currents
$j^{C}_{\pm,x}(\textbf{r},t)=j^{C}_{\pm,x}(\theta,\phi)$ can be obtained,
which satisfy the following relations,
\begin{align}
j^{C}_{-,x}(\theta,\phi)&=-j^{C}_{+,x}(\pi-\theta,\phi).
\end{align}
Because of this relation, the total charge current
$j^{C}_{x}(\theta,\phi)=j^{C}_{+,x}(\theta,\phi)+j^{C}_{-,x}(\theta,\phi)$
also satisfy the following condition,
\begin{align}\label{eqn:currentrelation1}
j^{C}_{x}(\theta,\phi)&=-j^{C}_{x}(\pi-\theta,\phi).
\end{align}
Therefore if the uniform component $N_{z}=\cos\theta(t)$ of $\hat{N}(\textbf{r},t)$
changes over the full polar angle $\theta(t)\in[0,\pi]$,
the net transferred charge due to the adiabatic charge current $j^{C}_{x}$
is zero, i.e.,
\begin{align}\label{eqn:netcharge}
Q_{\text{pumped}}=\int_{0}^{T}dt\int dy j^{C}_{x}=0.
\end{align}
On the other hand, for $\hat{N}^{\text{DW}}_{X}(\textbf{r},t)$ and $\hat{N}^{\text{DW}}_{Y}(\textbf{r},t)$,
it can be shown that
the adiabatic charge polarization is always zero for any given $\theta(t)$
after the integration over $\phi$, because of the following condition satisfied by $j^{C}_{x,\pm}(\theta,\phi)$,
\begin{align}\label{eqn:currentrelation2}
j^{C}_{x,+}(\theta,\phi)&=-j^{C}_{x,-}(\theta,\phi+\pi).
\end{align}
Therefore the quantized charge pumping is impossible for any orientation of $\textbf{m}_{\text{AF}}$.

The fact that the net transferred charge is zero can be understood
in terms of the spin gauge flux induced spin Hall effect.
If $|m_{\text{AF}}|\gg |t_{2}|$ where $\hat{N}_{+}$ and $\hat{N}_{-}$ are almost parallel,
the charge current density is proportional to $C_{S,+}+C_{S,-}$ as shown in Eq.~(\ref{eqn:currents_SHI})
and Eq.~(\ref{eqn:netcurrents_SHI}).
Since $C_{S,+}=-1$ and $C_{S,-}=1$ in this limit, the total charge current is zero
for given spin gauge flux.
The opposite sign of $C_{S,\pm}$ is the direct consequence of
the fact that the Dirac particles at the two nodal points have the opposite winding directions,
which is reflected in the opposite sign of the $k_{x}$ dispersions in Eq.~(\ref{eqn:Hplusminus}).~\cite{Graphene_review, Yang}
However, when $|m_{\text{AF}}|$ is finite, $\hat{N}_{+}$ and $\hat{N}_{-}$ are tilted with a small relative angle
due to the spin-orbit coupling. In this case, since the spin gauge fluxes generated by $\hat{N}_{\pm}$
are different, the local polarization current can be nonzero although
$C_{S,+}=-1$ and $C_{S,-}=1$ are still satisfied.
Therefore the finite partial polarization $\frac{dP_{x}}{d\theta}$ is essentially the result of
the spin anisotropy due to the spin-orbit coupling, not of a topological origin.
On the other hand, if $|m_{\text{AF}}|\ll |t_{2}|$ corresponding to SHI phase, the solid angle subtended by $\hat{N}_{\pm}$
is very tiny even though $\hat{N}$ covers the full spherical solid angle,
which leads to vanishingly small charge polarization.
The quantized pumping is also impossible in SHI case.


\subsection{\label{sec:Skyrmion_SHI} Skyrmion charge}

The skyrmion defects of the collinear magnetic phases satisfying
$\textbf{m}_{\text{AF}} \parallel\textbf{t}_{2}$ or $\textbf{m}_{\text{AF}} \perp \textbf{t}_{2}$
can also be described by using the spin configurations
$\hat{N}^{\text{DW}}_{X,Y,Z}$ in Eq.~(\ref{eqn:XYZspiral}).
For the description of a skyrmion, we replace $\theta(t)$ by $\theta(|\textbf{r}|)$ satisfying the boundary conditions
$\theta(|\textbf{r}|=0)=\pi$ and $\theta(|\textbf{r}|\rightarrow\infty)=0$
and adopt $\phi(\textbf{r})=\tan^{-1}(\frac{y}{x})$ with $\textbf{r}=(x,y)$.
In the case of $\textbf{m}_{\text{AF}} \parallel\textbf{t}_{2}$,
a skyrmion defect can be described by using $\hat{N}^{\text{DW}}_{Z}$.
As Eq.~(\ref{eqn:partialpolarization}), Eq.~(\ref{eqn:currentrelation1}) and Fig.~\ref{fig:DW_polarization} (b)
show, the adiabatically induced charge of a skyrmion, which is equivalent to the partial polarization of a domain wall,
is anti-symmetrically distributed with respect to $(\theta-\pi/2)$.
Therefore the integration of the induced charge adds up to zero,
which shows that the skyrmion is not charged.
On the other hand, when $\textbf{m}_{\text{AF}} \perp \textbf{t}_{2}$,
a skyrmion can be described by $\hat{N}^{\text{DW}}_{X}$ or $\hat{N}^{\text{DW}}_{Y}$.
In this case, as shown in Eq.~(\ref{eqn:currentrelation2}),
the induced charge is zero for any $\theta$ value.
Therefore the skyrmion is not charged.

To sum up, the skyrmions in CMIs derived from SHI
do not carry the charge quantum number at least for
$|m_{\text{AF}}|<||t_{2}|-|\mu_{S}||$ or $|m_{\text{AF}}|>||t_{2}|+|\mu_{S}||$,
which corresponds to SHI or SDW phases, respectively.

\subsection{\label{sec:Meron_SHI} Meron charge}

The anti-symmetric distribution of the induced charge with respect to $(\theta-\pi/2)$, shown in
Fig.~\ref{fig:DW_polarization} (b), implies that a meron-type defect in which $\theta$ changes over $\theta\in[0,\pi/2]$
can be charged although a skyrmion is chargeless.
For $\textbf{m}_{\text{AF}} \parallel\textbf{t}_{2}$,
a meron can be described by $\hat{N}^{\text{DW}}_{Z}$ with the boundary condition
given by $\theta(|\textbf{r}|=0)=\pi/2$ and $\theta(|\textbf{r}|\rightarrow\infty)=0$.
Explicitly, from the adiabatic charge current in Eq.~(\ref{eqn:currents_SHI}),
the meron charge $Q_{\text{meron}}$ is given by
\begin{align}
Q_{\text{meron}}=1-\frac{t_{2}}{\sqrt{m_{\text{AF}}^{2}+t_{2}^{2}}}\approx \frac{m_{\text{AF}}^{2}}{2t_{2}^{2}},
\end{align}
for SHI phase satisfying $t_{2}\gg m_{\text{AF}}>0$. Similarly for SDW phase which satisfy
$m_{\text{AF}}\gg t_{2}>0$,
\begin{align}
Q_{\text{meron}}=-\frac{t_{2}}{\sqrt{m_{\text{AF}}^{2}+t_{2}^{2}}}\approx -\frac{t_{2}}{m_{\text{AF}}}.
\end{align}
Therefore a meron defect has a finite charge, however, which is not quantized.

A meron defect is especially important for the spin ordering with easy-plane anisotropy.
Considering a planar antiferromagnetic spin ordering $\textbf{m}_{\text{AF}}=(m_{x},m_{y},0)$,
the low energy Hamiltonian can be written as
\begin{align}\label{eqn:Hplusminus_meron1}
H_{\text{eff},\pm}=&\int d^{2}r \psi^{\dag}_{\pm}(\textbf{r},t)[\pm(-i\partial_{x})\tau_{x}+(-i\partial_{y})\tau_{y}
+\mu_{S}\tau_{z}
\nonumber\\
&+\tau_{z}\{\pm t_{2}\sigma_{z}
+m_{x}\sigma_{x}+m_{y}\sigma_{y}\}]\psi_{\pm}(\textbf{r},t),
\end{align}
where $\pm$ indicates the two nodes at $\textbf{k}=\pm \textbf{K}_{1}$.
After a transformation of $\psi_{-}\rightarrow\tau_{y}\sigma_{z}\psi_{-}$,
the full Hamiltonian $H_{\text{eff}}=H_{\text{eff},+}+H_{\text{eff},-}$ can be written as
\begin{align}\label{eqn:Hplusminus_meron2}
H_{\text{eff}}=&\int d^{2}r \psi^{\dag}(\textbf{r},t)[(-i\partial_{x})\tau_{x}+(-i\partial_{y})\tau_{y}
+\mu_{S}\tau_{z}\nu_{z}
\nonumber\\
&+\tau_{z}\{t_{2}\sigma_{z}
+m_{x}\sigma_{x}+m_{y}\sigma_{y}\}]\psi(\textbf{r},t),
\end{align}
where $\psi^{\dag}=(\psi^{\dag}_{+},\psi^{\dag}_{-})$.
It is interesting to notice that this Hamiltonian has similar structure as the
effective Hamiltonian of CMI derived from AHI in Eq.~(\ref{eqn:effH1}).
Now we define a three component unit vector $\hat{N}=\frac{1}{\sqrt{m_{x}^{2}+m_{y}^{2}+t_{2}^{2}}}(m_{x},m_{y},t_{2})$
and compute the charge current induced by the adiabatic variation of $\hat{N}$ using the formulation in Sec.~\ref{sec:AHI}.
A meron can be represented by
\begin{align}
\hat{N}_{\text{meron}}=(\cos\alpha(r)\cos\phi,\cos\alpha(r)\sin\phi,\sin\alpha(r)),
\end{align}
which is under the constraint of $\alpha(r=0)=\frac{\pi}{2}$ and
$\sin[\alpha(r\rightarrow\infty)]=\frac{t_{2}}{\sqrt{m^{2}+t_{2}^{2}}}$.
Here $m^{2}=m_{x}^{2}+m_{y}^{2}$, $\phi(\textbf{r})=\tan^{-1}(\frac{y}{x})$, and $|\textbf{r}|=r=\sqrt{x^{2}+y^{2}}$.
If $|\mu_{\text{S}}|>\sqrt{m^{2}+t_{2}^{2}}$, the induced charge is zero.
On the other hand, when $|\mu_{\text{S}}|<\sqrt{m^{2}+t_{2}^{2}}$, the total charge induced by a meron defect
is given by
\begin{align}
Q_{\text{meron}}
&=\frac{C_{S}}{4\pi}\int dxdy \hat{N}_{\text{meron}}\cdot[(\partial_{x}\hat{N}_{\text{meron}})\times(\partial_{y}\hat{N}_{\text{meron}})],
\nonumber\\
&=\frac{t_{2}}{\sqrt{m^{2}+t_{2}^{2}}}-1,
\end{align}
which shows that a meron defect is charged but the net charge is not quantized.
$Q_{\text{meron}}\approx\frac{-m^{2}}{2t_{2}^{2}}$ for $t_{2}\gg m$ while
$Q_{\text{meron}}$ approaches -1 for $m\gg t_{2}$.
The meron charge of the easy-plane Neel phase derived from SHI is also discussed
by the recent work in Ref.~\onlinecite{Lee_meron}, in which
the same conclusion is obtained.

\section{\label{sec:Semiclassical} Second Chern number and topological contribution to the inhomogeneity induced adiabatic currents}
\subsection{\label{sec:Semiclassical_a} Semiclassical gradient expansion approach for adiabatic currents}

In this section, we demonstrate that
the charge and spin polarizations induced by the adiabatic variation
of the inhomogeneous spin order parameter $\hat{N(\textbf{r},t)}$
can be understood as a generic topological response of inhomogeneous systems
under adiabatic temporal variations.
In particular, it is shown that the topological
nature of the adiabatic current of inhomogeneous systems is endowed with
the second Chern number. It turns out that the existence of the second Chern
number is the topological origin of the fact that the spin and charge polarization quantums
induced by the inhomogeneous spin order parameter $\hat{N(\textbf{r},t)}$
are determined by the product of two topological invariants,
which are defined in the position and momentum spaces, respectively,
as shown in Eq.~(\ref{eqn:chargequantum}) and Eq.~(\ref{eqn:spinquantum}).

For simplicity, we assume that the system has translational invariance
along the $x$-direction and the spatial inhomogeneity is developed
along the $y$-direction. The total charge current along the $x$-direction, $J_{x}$
is given by
\begin{align}
J_{x}&=\frac{\delta}{\delta A_{x}}i\text{Trln}G
\nonumber\\
&=-ie\int\frac{dk_{x}}{2\pi}\int dy\int dt\text{tr}\langle t,y|\hat{G}(k_{x})\partial_{k_{x}}\hat{G}^{-1}(k_{x})|t,y\rangle,
\end{align}
where the symbol $\text{tr}$ indicates the trace
over the discrete indices while $\text{Tr}$ includes the trace over
discrete indices and integration over the momentum $k_{x}$
and the position $y$ and time $t$.
Since the electrons are coupled to the time-dependent
inhomogeneous order parameter $\hat{N}(y,t)$,
the Green's function, which is diagonal in $k_{x}$, has
off-diagonal components in the $(y,t)$ space, i.e.,
$\langle y_{1},t_{1}|\hat{G}(k_{x})|y_{2},t_{2}\rangle=G(k_{x};y_{1},t_{1};y_{2},t_{2})$.

If $\hat{N}(y,t)$ changes slowly in the position and time spaces, namely,
if $L_{y}\gg 1/E_{g}$ and $1/T \ll E_{g}$ are satisfied,
the position and time dependence of the Green's function $G$
can be treated by the gradient expansion method.
Here $L_{y}$ ($T$) is the length (time) scale over which $\hat{N}(y,t)$ varies and $E_{g}$ is the bulk energy gap.
To perform the systematic gradient expansion,
we first consider the Wigner transformation
of $G(k_{x};y_{1},t_{1};y_{2},t_{2})$,
i.e. the Fourier transformation with respect to the relative coordinates $y_{1}-y_{2}$ and $t_{1}-t_{2}$,
which can be written as~\cite{GE_Rammer, GE_Gurarie, GE_Baraff}
\begin{align}\label{eqn:WTG}
&\widetilde{G}(k_{x};k_{y},\frac{y_{1}+y_{2}}{2};\omega,\frac{t_{1}+t_{2}}{2})
\nonumber\\
&\equiv\int d(t_{1}-t_{2})d(y_{1}-y_{2}) e^{-i[k_{y}(y_{1}-y_{2})-\omega(t_{1}-t_{2})]}
\nonumber\\
&\qquad\times G(k_{x};y_{2},t_{1};y_{2},t_{2}).
\end{align}
As shown in detail in the Appendix~\ref{sec:GradientExpansion}, the key ingredient
of the gradient expansion is that the Wigner transformed
Green's function $\widetilde{G}$ can be expanded order by order
in powers of the spatial gradient of
the semi-classical Green's function $\mathcal{G}(k_{x};k_{y},y;\omega,t)=[\omega-\mathcal{H}_{\text{sc}}(k_{x};k_{y},y;\omega,t)]^{-1}$.~\cite{GE_Baraff, Volovik_Book}
In the semi-classical Hamiltonian $\mathcal{H}_{\text{sc}}(k_{x};k_{y},y;\omega,t)$, the conjugate variables
$(y,\hat{k}_{y})$ and $(t,\hat{\omega})$ are treated as independent real numbers $(y,k_{y})$ and $(t,\omega)$.
Here $\hat{k}_{y}=-i\frac{\partial}{\partial_{y}}$ and $\hat{\omega}=i\frac{\partial}{\partial_{t}}$.
Therefore the gradient expansion is equivalent to the semi-classical expansion.

The leading order term of the adiabatic current induced by the spatial inhomogeneity
can be written as
\begin{align}
J_{x}=J_{x,\text{topological}}+J_{x,\text{perturb}},
\end{align}
in which
\begin{align}\label{eqn:Jtopology1}
J_{x,\text{topological}}
=&-\frac{ie}{60}\int\frac{d\omega}{2\pi}\int_{T^{2}} \frac{d^{2}\textbf{k}}{(2\pi)^{2}}\int dy dt
\nonumber\\
&\times\varepsilon_{\nu_{1}\nu_{2}\nu_{3}\nu_{4}\nu_{5}}
\text{tr}[\mathcal{G}(\partial_{\nu_{1}}\mathcal{G}^{-1})\cdot\cdot\cdot\mathcal{G}(\partial_{\nu_{5}}\mathcal{G}^{-1})],
\end{align}
where the indices $(\nu_{1},\nu_{2},\nu_{3},\nu_{4},\nu_{5})$ run over $(\omega,k_{x},k_{y},y,t)$.
Here $T^{2}$ indicates the 2-torus made of two dimensional Brillouin zone.
If the $(y,t)$ dependence of the Hamiltonian
occurs only through the coupling to $\hat{N}(y,t)$,
$J_{x,\text{topological}}$ can be rewritten using the spherical coordinates $(\theta,\phi)$
for the unit vector $\hat{N}(y,t)=\hat{N}(\theta(y,t),\phi(y,t))$
in the following way,
\begin{align}\label{eqn:Jtopology2}
J_{x,\text{topological}}
=&-\frac{ie}{60}\int\frac{d\omega}{2\pi}\int_{T^{2}} \frac{d^{2}\textbf{k}}{(2\pi)^{2}}\int_{S^2} d\theta d\phi
\nonumber\\
&\times\varepsilon_{\mu_{1}\mu_{2}\mu_{3}\mu_{4}\mu_{5}}
\text{tr}[\mathcal{G}(\partial_{\mu_{1}}\mathcal{G}^{-1})\cdot\cdot\cdot\mathcal{G}(\partial_{\mu_{5}}\mathcal{G}^{-1})]
\nonumber\\
=&eC^{(2)}_{\text{Chern}}
\end{align}
where the indices $(\mu_{1},\mu_{2},\mu_{3},\mu_{4},\mu_{5})$ run over $(\omega,k_{x},k_{y},\theta,\phi)$.
When $\hat{N}(y,t)$ fully covers the unit sphere during the adiabatic variation,
the total adiabatic current is quantized with the magnitude of the quantum
determined by the second Chern number $C^{(2)}_{\text{Chern}}$.
All the other non-topological terms, whose contribution
would vanish for cyclic evolutions of $\hat{N}$, are
included in $J_{x,\text{perturb}}$.
The explicit expressions of $J_{x,\text{perturb}}$ are given in
Eq.~(\ref{eqn:J2A}), Eq.~(\ref{eqn:J2B}), and Eq.~(\ref{eqn:J2C}).

The charge density induced by the spatial inhomogeneity can also be derived
in a similar way from the semiclassical gradient expansion approach.
Starting from the expression of the integrated charge density which is given by
\begin{align}
J_{0}&=\frac{\delta}{\delta A_{0}}i\text{Trln}G
\nonumber\\
&=-ie\int\frac{d\omega}{2\pi}\int dx\int dy\text{tr}\langle x,y|\hat{G}(\omega)\partial_{\omega}\hat{G}^{-1}(\omega)|x,y\rangle,
\end{align}
the inhomogeneity induced charge can be obtained from the expansion of the Wigner transformed Green's
function in the powers of the
spatial gradients $\partial_{x}$ and $\partial_{y}$.
The lowest order contribution to the adiabatic charge induced by a skyrmion defect
appears in the second order terms which contains $\partial_{x}$ and $\partial_{y}$ at the same time.
Explicitly, it is given by
\begin{align}\label{eqn:J0topology}
Q_{\text{skyrmion}}
=&-\frac{ie}{60}\int\frac{d\omega}{2\pi}\int_{T^{2}} \frac{d^{2}\textbf{k}}{(2\pi)^{2}}\int_{S^2} d\theta d\phi
\nonumber\\
&\times\varepsilon_{\mu_{1}\mu_{2}\mu_{3}\mu_{4}\mu_{5}}
\text{tr}[\mathcal{G}(\partial_{\mu_{1}}\mathcal{G}^{-1})\cdot\cdot\cdot\mathcal{G}(\partial_{\mu_{5}}\mathcal{G}^{-1})]
\end{align}
where the indices $(\mu_{1},\mu_{2},\mu_{3},\mu_{4},\mu_{5})$ run over $(\omega,k_{x},k_{y},\theta,\phi)$.
Here $\theta$ and $\phi$ represents the spherical coordinates
for the three component unit vector $\hat{N}(x,y)=\hat{N}(\theta(x,y),\phi(x,y))$
describing a skyrmion configuration.
The same expression for the skyrmion induced charge was obtained in the recent work
by Santos et al. in Ref.~\onlinecite{Santos}.
It is interesting to notice that $Q_{\text{skyrmion}}$ can be obtained simply by replacing
the time coordinate $t$ by the spatial coordinate $x$
from the expression of the total adiabatic charge current $J_{x,\text{topological}}$
in Eq.~(\ref{eqn:Jtopology2}).

Now we show that the appearance of the second Chern number in $J_{x,\text{topological}}$
is the topological origin of the fact that the quantum of the pumped charge in CMIs
is given by the product of two topological invariants as shown in Eq.~(\ref{eqn:chargequantum}).
Let us consider the following semi-classical Green's function $\mathcal{G}$ which
describes the low energy excitations of CMIs,~\cite{Footnote_SCGreen}
\begin{align}\label{eqn:SMG_4by4}
\mathcal{G}^{-1}=\omega-f_{1}(\textbf{k})\tau_{x}-f_{2}(\textbf{k})\tau_{y}
-m\tau_{z}\hat{N}(y,t)\cdot\vec{\sigma},
\end{align}
where $\tau_{x,y,z}$ and $\sigma_{x,y,z}$ are Pauli matrices
and $f_{1,2}(\textbf{k})$ are functions which depend only on the momenta $\textbf{k}$.
Let us consider the property of the following term,
\begin{align}
&I_{N_{a}N_{b}k_{1}k_{2}\omega}
\nonumber\\
&\equiv\text{tr}[\mathcal{G}(\partial_{N_{a}}\mathcal{G}^{-1})\mathcal{G}(\partial_{N_{b}}\mathcal{G}^{-1})
\mathcal{G}(\partial_{k_{1}}\mathcal{G}^{-1})\mathcal{G}(\partial_{k_{2}}\mathcal{G}^{-1})
\mathcal{G}(\partial_{\omega}\mathcal{G}^{-1})],
\end{align}
where $N_{a}$ is a component of the unit vector $\hat{N}=(N_{x},N_{y},N_{z})$.
We introduce a local unitary transformation $U$ that satisfies
$U^{\dag}(\vec{\sigma}\cdot\hat{N})U=\sigma_{z}$.
The semi-classical Green's function in the rotated frame,
which is defined as $\mathcal{G}_{D}\equiv U^{\dag}\mathcal{G}U$,
is diagonal in the spin $\vec{\sigma}$ space.
After inserting the identity $U^{\dag}U=1$ between $\mathcal{G}$ and $\mathcal{G}^{-1}$,
$I_{N_{a}N_{b}k_{1}k_{2}\omega}$ transforms as
\begin{align}
I_{N_{a}N_{b}k_{1}k_{2}\omega}
&=\text{tr}[U^{\dag}\mathcal{G}(\partial_{N_{a}}\mathcal{G}^{-1})\mathcal{G}(\partial_{N_{b}}\mathcal{G}^{-1})U
\nonumber\\
&\qquad\times\mathcal{G}_{D}(\partial_{k_{1}}\mathcal{G}_{D}^{-1})\mathcal{G}_{D}(\partial_{k_{2}}\mathcal{G}_{D}^{-1})
\mathcal{G}_{D}(\partial_{\omega}\mathcal{G}_{D}^{-1})].
\end{align}
Using $\mathcal{G}$ from Eq.~(\ref{eqn:SMG_4by4}), it can be shown that
\begin{align}
U^{\dag}\mathcal{G}(\partial_{N_{a}}\mathcal{G}^{-1})\mathcal{G}(\partial_{N_{b}}\mathcal{G}^{-1})U
=\frac{im^{2}}{\omega^{2}-E^{2}}\varepsilon_{abc}U^{\dag}\sigma_{c}U.
\end{align}
Taking into account the fact that $\mathcal{G}_{D}$ is spin-diagonal,
it can be shown that
\begin{align}
&I_{N_{a}N_{b}k_{1}k_{2}\omega}
\nonumber\\
&=\frac{im^{2}}{\omega^{2}-E^{2}}\varepsilon_{abc}N_{c}
\text{tr}[\sigma_{z}\mathcal{G}_{D}(\partial_{k_{1}}\mathcal{G}_{D}^{-1})\mathcal{G}_{D}(\partial_{k_{2}}\mathcal{G}_{D}^{-1})
\mathcal{G}_{D}(\partial_{\omega}\mathcal{G}_{D}^{-1})],
\end{align}
where $E^{2}=f_{1}^{2}+f_{2}^{2}+m^{2}$. Then
\begin{align}
I_{ytk_{1}k_{2}\omega}&=\frac{\partial N_{a}}{\partial y}\frac{\partial N_{b}}{\partial t}I_{N_{a}N_{b}k_{1}k_{2}\omega}
\nonumber\\
&=\frac{im^{2}}{\omega^{2}-E^{2}}\hat{N}\cdot\big((\partial_{y}\hat{N})\times(\partial_{t}\hat{N})\big)
\nonumber\\
&\times\text{tr}[\sigma_{z}\mathcal{G}_{D}(\partial_{k_{1}}\mathcal{G}_{D}^{-1})\mathcal{G}_{D}(\partial_{k_{2}}\mathcal{G}_{D}^{-1})
\mathcal{G}_{D}(\partial_{\omega}\mathcal{G}_{D}^{-1})].
\end{align}
Interestingly, the inhomogeneous part that depends on $(y,t)$ and
the homogeneous part that is written in term of $\mathcal{G}_{D}$
in the $(\omega,k_{x},k_{y})$ space are completely separated.
Applying this result to Eq.~(\ref{eqn:Jtopology1}),
$J_{x,\text{topological}}$ can be written as
\begin{align}
J_{x,\text{topological}}=\frac{C_{S}}{4\pi}\int dydt \hat{N}\cdot\big[(\partial_{y}\hat{N})\times(\partial_{t}\hat{N})\big],
\end{align}
where the spin Chern number $C_{S}$ is defined in terms
of $\mathcal{G}_{D}$ using Eq.~(\ref{eqn:ChernNumbers})
after replacing $G_{0}$ by $\mathcal{G}_{D}$.
It is to be noted that the total charge current $J_{x,\text{topological}}$ induced
by the adiabatic change of the spatial inhomogeneity
is equal to the net transferred charge, which is shown in Eq.~(\ref{eqn:chargequantum}).
The existence of the topological contribution to the inhomogeneity
induced charge current characterized by the second Chern number
is the origin of the quantized charge pumping with the quantum of the pumped charge
expressed as the product of two distinct topological invariants.

Following the similar line of reasoning, the inhomogeneity induced
total spin current can be obtained in the following way,
\begin{align}\label{eqn:JStopology1}
&J^{S}_{x,\text{topological}}
=-\frac{ie}{60}\int\frac{d\omega}{2\pi}\int_{T^{2}} \frac{d^{2}\textbf{k}}{(2\pi)^{2}}\int dy dt
\nonumber\\
&\quad\times\varepsilon_{\nu_{1}\nu_{2}\nu_{3}\nu_{4}\nu_{5}}
\text{tr}[\big(\frac{\hbar}{2}\vec{\sigma}\cdot\hat{N}(y,t)\big)\mathcal{G}(\partial_{\nu_{1}}\mathcal{G}^{-1})
\cdot\cdot\cdot\mathcal{G}(\partial_{\nu_{5}}\mathcal{G}^{-1})],
\end{align}
For the semi-classical Green's function in Eq.~(\ref{eqn:SMG_4by4})
it is straightforward to show that,
\begin{align}
J^{S}_{x,\text{topological}}=\frac{\hbar}{2}\frac{C_{C}}{4\pi}\int dydt \hat{N}\cdot\big[(\partial_{y}\hat{N})\times(\partial_{t}\hat{N})\big],
\end{align}
where $C_{C}$ is the charge Chern number defined in terms of $\mathcal{G}$.
It is interesting to notice the equivalence of $J^{S}_{x,\text{topological}}$ above
and the total transferred spin $S_{\text{transfer}}$ in Eq.~(\ref{eqn:spinquantum}).

\begin{figure}[t]
\centering
\includegraphics[width=8.5 cm]{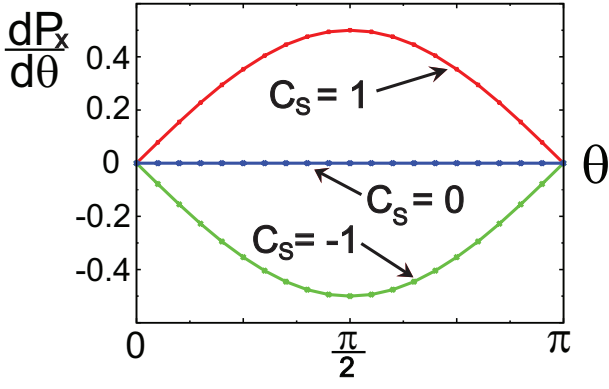}
\caption{(Color online)
Inhomogeneity induced adiabatic polarization ($P_{x}$)
for a CMI derived from AHI. A spiral spin ordering with the
modulation period of $L_{y}$ along the $y$-direction is assumed.
$\frac{dP_{x}}{d\theta}$
is plotted in the unit of $e$ using Eq.~(\ref{eqn:2ndChern}).
For $t_{2}$=2, $\mu_{S}$=1,
$\frac{dP_{x}}{d\theta}$ is computed for $m_{\text{AF}}$=-2, 0.5, 2 and 4
which correspond to a TCMI ($C_{S}=-1$),
an AHI ($C_{S}=0$), another
TCMI ($C_{S}=1$)
and a SDW ($C_{S}=0$), respectively.
Notice that the polarization is solely determined by
the spin Chern number $C_{S}$.
} \label{fig:polarization}
\end{figure}

Recently, the topological contribution to the inhomogeneity induced
adiabatic charge polarization was investigated by considering
the Berry phase approach to the charge polarization.~\cite{Vanderbilt,Resta,Oritz}
Using the semiclassical wave packet dynamics formalism, Xiao et al.,~\cite{Niu, Niu_review}
have derived the general expression of the adiabatic polarization induced
by spatial inhomogeneity. For a 2D multi-band system in which the smooth spatial modulation exists
along the $\beta$ direction ($\beta=x, y$)
with the modulation period of $L_{\beta}$, the net charge polarization
along the $\alpha$ direction perpendicular to the modulation direction $\beta$
is given by~\cite{Niu, Niu_review, Essin},

\begin{align}\label{eqn:2ndChern}
&P^{(2)}_{\alpha}
=\int dr_{\beta}\int_{0}^{T} dt
j_{\alpha}(\textbf{r},t),
\nonumber\\
&\qquad=\frac{e}{8} \int dr_{\beta}\int_{\lambda(0)}^{\lambda(T)} d\lambda\int_{\text{BZ}}\frac{d\textbf{k}}{(2\pi)^d}
\epsilon_{ijkl}\text{Tr}[F_{ij}F_{kl}],
\end{align}
where $j_{\alpha}(\textbf{r},t)$ is the inhomogeneity induced
adiabatic current as the parameter $\lambda(t)$ evolves from
$\lambda(0)$ to $\lambda(T)$.
Here the indices $i, j, k, l$ run over $(k_{\alpha},k_{\beta},r_{\beta},\lambda)$ and
the Berry curvature $F_{ij}$ satisfies $F_{ij}=\partial_{i}A_{j}-\partial_{j}A_{i}-i[A_{i},A_{j}]$
with the Berry connection $A_{j}^{\mu\nu}=\langle \Phi_{\mu}|i\partial_{j}|\Phi_{\nu}\rangle$.
Here $|\Phi_{\nu}\rangle$ is the periodic part of the Bloch wave function
for the occupied band $\nu$.
The equivalence of the $P^{(2)}_{x}$ in Eq.~(\ref{eqn:2ndChern}) and
$J_{x,\text{topological}}$ in Eq.~(\ref{eqn:Jtopology1}) can be proved
by considering the topological invariance of the second Chern number
when the order parameter $\hat{N}(y,t)$ spans a closed manifold.
The details of the proof is given in Ref.~\onlinecite{Qi}.
Therefore the inhomogeneity induced charge polarization can be computed, for a general (lattice) Hamiltonian,
using either Eq.~(\ref{eqn:Jtopology1}) or Eq.~(\ref{eqn:2ndChern}).

\begin{figure}[t]
\centering
\includegraphics[width=8.5 cm]{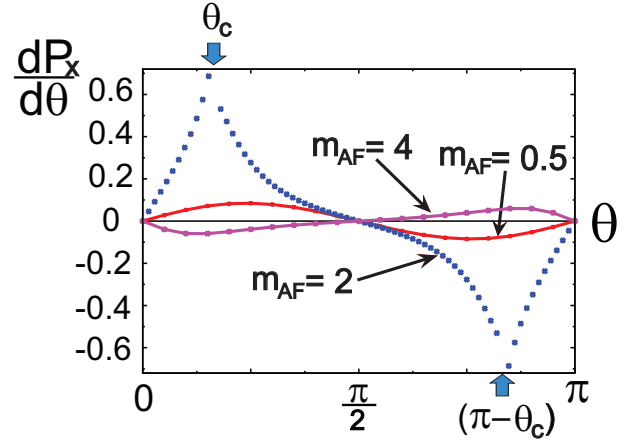}
\caption{(Color online)
Inhomogeneity induced adiabatic polarization ($P_{x}$)
for a CMI derived from SHI.
A spiral spin ordering with the
modulation period of $L_{y}$ along the $y$-direction is assumed.
$\frac{dP_{x}}{d\theta}$
is plotted in the unit of $e$ using Eq.~(\ref{eqn:2ndChern}).
For $t_{2}$=2, $\mu_{S}$=1,
$\frac{dP_{x}}{d\theta}$ is computed for $m_{\text{AF}}$=0.5, 2 and 4,
which are relevant to SHI, TCMI, SDW phases in Fig.~\ref{fig:phasediagram_SHI} (c),
respectively, in
the collinear limit.
$\theta=\theta_{c}$ and $\theta=(\pi-\theta_{c})$ indicate the locations where
the singularity in $\frac{dP_{x}}{d\theta}$ occurs for $m_{\text{AF}}$=2.
} \label{fig:polarizationfromSHI}
\end{figure}

\subsection{\label{sec:Semiclassical_b} Adiabatic polarization and skyrmion charge for CMI derived from AHI}

To demonstrate the adiabatic polarization induced by inhomogeneous ferrimagnetic ground states,
we consider a spiral (or conical) spin ordering pattern with the modulation period $L_{y}=2\pi/Q$.
The periodic spin modulation along the $y$ direction can also be represented by
use of $\hat{N}^{\text{DW}}_{X,Y,Z}$ in Eq.~(\ref{eqn:XYZspiral})
taking $\phi(y)=Qy$.
The spatial modulation of the spin ordering is described in Fig.~\ref{fig:spiral}.
The polarization of the full lattice Hamiltonian $\hat{H}_{\text{full}}$ in Eq.~(\ref{eqn:fullHamiltonian})
is numerically computed from the second Chern form given in Eq.~(\ref{eqn:2ndChern}).
Since the modulated spin ordering spreads over the whole system,
$S_{z}$ is not conserved any more. Therefore we focus on the
charge polarization induced by the inhomogeneous spin ordering.

Let us first consider the CMI derived from AHI.
In Fig.~\ref{fig:polarization}, we plot the derivative of average polarization $\frac{dP_{x}}{d\theta}$
per unit modulation period $L_{y}$
as a function of the polar angle $\theta(\lambda)$ which plays the role of the adiabatic parameter.
For $t_{2}$=2, $\mu_{S}$=1,
the polarization is computed for $m_{\text{AF}}$=-2, 0.5, 2 and 4
which, in the collinear limit, correspond to a TCMI ($C_{S}=-1$),
an AHI ($C_{S}=0$), another
TCMI ($C_{S}=1$)
and a SDW ($C_{S}=0$), respectively, following
the phase diagram in Fig.~\ref{fig:phasediagram_AHI} (b).
It is worth to be noted that the magnitude and sign of the polarization are solely determined
by the spin Chern number $C_{S}$.
Moreover, as $\theta$ evolves from 0 to $\pi$,
the average polarization per unit modulation period $L_{y}$ is quantized satisfying
$\int_{0}^{\pi} d\theta \frac{dP_{x}}{d\theta}=C_{S}$.
Therefore quantized charge pumping is possible
in CMI derived from AHI, consistent with the conclusion in Sec.~\ref{sec:TCMI_AHI}.
At the same time, this means that a skyrmion defect made of the staggered spin $\hat{N}$
has the charge quantum number $C_{S}$.
The polarization (or skyrmion charge) is also computed including
the ferromagnetic component of the spin order parameter $m_{\text{F}}$
considering the general ferrimagnetic ordering with a net ferromagnetic moment.
We have confirmed that $m_{\text{F}}$
does not affect the polarization as long as
the bulk band gap is not closed by $m_{\text{F}}$.

\subsection{\label{sec:Semiclassical_c} Adiabatic polarization and skyrmion charge for CMI derived from SHI}

In the case of the CMI derived from SHI, $\frac{dP_{x}}{d\theta}$ depends
on the orientation of the uniform component $\cos\theta$ of the rotating spins
relative to the anisotropy direction $\textbf{t}_{2}$ inherent to the SHI.
In the case of the spiral ordering described by $\hat{N}^{\text{DW}}_{X}$ or $\hat{N}^{\text{DW}}_{Y}$ in Eq.~(\ref{eqn:XYZspiral}),
in which the uniform component aligns perpendicular to the spin anisotropy direction,
$\frac{dP_{x}}{d\theta}=0$ always.
Therefore the net pumped charge is zero in this case.

On the other hand, for the spiral ordering described by $\hat{N}^{\text{DW}}_{Z}$ in Eq.~(\ref{eqn:XYZspiral}),
which corresponds to the transverse conical phase with the uniform spin component parallel to $\textbf{t}_{2}$,
$\frac{dP_{x}}{d\theta}$ is plotted in Fig.~\ref{fig:polarizationfromSHI}.
For $t_{2}$=2, $\mu_{S}$=1,
the polarization is computed for $m_{\text{AF}}$=0.5, 2 and 4,
which are relevant to SHI, TCMI, SDW phases in Fig.~\ref{fig:phasediagram_SHI} (c), respectively, in
the collinear limit.
$\frac{dP_{x}}{d\theta}$ is odd function with respect to $\theta=\frac{\pi}{2}$,
which is consistent with Eq.~(\ref{eqn:currentrelation1}).
Because of this, as $\theta$ evolves from 0 to $\pi$,
the net polarization is zero, i.e.,
$\int_{0}^{\pi} d\theta \frac{dP_{x}}{d\theta}=0$.
Therefore the quantized charge pumping is impossible
via the adiabatic rotation of the uniform component of the spiral spins
in this case.
At the same time, it means that the skyrmion defect has no electric charge
independent of $|m_{\text{AF}}|$ in the case of CMI derived from SHI.

The CMIs with $m_{\text{AF}}$=2 satisfy the condition of $|t_{2}-\mu_{S}|<|m_{\text{AF}}|<|t_{2}+\mu_{S}|$,
in which the adiabatic polarization could not be properly described by the effective
gauge field approach used in Sec.~\ref{sec:TCMI_SHI}.
In this case, $\frac{dP_{x}}{d\theta}$ shows strong enhancement
near $\theta=\theta_{c}$ and $\pi-\theta_{c}$ as shown in Fig.~\ref{fig:polarizationfromSHI}.
At the singular points of $\theta=\theta_{c}$ and $\pi-\theta_{c}$,
$\frac{dP_{x}}{d\theta}$ is not well-defined.
Comparing the magnitude of $\frac{dP_{x}}{d\theta}$ near $\theta=\theta_{c}$ or $\theta=\pi-\theta_{c}$,
the partial
polarization for $m_{\text{AF}}$=2 is an order of magnitude larger than those for $m_{\text{AF}}=0.5$
or $m_{\text{AF}}=4$.
The origin of this enhancement can be understood in the following way.
The semiclassical Hamiltonian describing the low energy electrons near $\textbf{k}=\pm \textbf{K}_{1}$
in CMI derived from SHI is given by
\begin{align}\label{eqn:SemiclassicH}
&\mathcal{H}_{\text{sc}}(\textbf{k},\textbf{r})
\nonumber\\
&=k_{x}\tau_{x}\nu_{z}+k_{y}\tau_{y}
+\mu_{S}\tau_{z}+t_{2}\tau_{z}\nu_{z}\sigma_{z}
+m_{\text{AF}}\tau_{z}\hat{N}(\textbf{r},t)\cdot\vec{\sigma}.
\end{align}
Using the semiclassical Green's function $\mathcal{G}^{-1}=\omega-\mathcal{H}_{sc}$,
the local charge Chern number can be defined in the following way,
\begin{align}
C_{C}(\textbf{r},t)=&\frac{\epsilon^{\mu\nu\lambda}}{24\pi^2}\int d^{3}k
\text{Tr}[\mathcal{G}\frac{\partial \mathcal{G}^{-1}}{\partial k_{\mu}}\mathcal{G}\frac{\partial \mathcal{G}^{-1}}{\partial k_{\nu}}\mathcal{G}
\frac{\partial \mathcal{G}^{-1}}{\partial k_{\lambda}}],
\nonumber\\
=&C_{+}(\textbf{r},t)+C_{-}(\textbf{r},t),
\end{align}
in which
\begin{align}
C_{\pm}(\textbf{r},t)=
(\mp)\frac{1}{2}\Big[&\text{sign}\Big(\mu_{S}+\sqrt{m^{2}_{\text{AF}}+t^{2}_{2}\pm 2m_{\text{AF}}t_{2}N_{z}}\Big)
\nonumber\\
&+\text{sign}\Big(\mu_{S}-\sqrt{m^{2}_{\text{AF}}+t^{2}_{2}\pm 2m_{\text{AF}}t_{2}N_{z}}\Big)\Big].
\end{align}
In contrast to the cases of $|m_{\text{AF}}|>|t_{2}+\mu_{S}|$ or $|m_{\text{AF}}|<|t_{2}-\mu_{S}|$
in which $C_{C}(\textbf{r},t)=0$ for any $\textbf{r}$ and $t$, when
$|t_{2}-\mu_{S}|<|m_{\text{AF}}|<|t_{2}+\mu_{S}|$,
$C_{C}(\textbf{r},t)$ changes discontinuously in the $(\textbf{r},t)$ space because
the energy spectrum of the semiclassical Hamiltonian $\mathcal{H}_{sc}$ develops band touching points.
At $\theta=\theta_{c}$ and $\pi-\theta_{c}$ where $\frac{dP_{x}}{d\theta}$ develops singularities, $\mathcal{H}_{sc}$
possesses gapless points at $\textbf{k}=\textbf{K}_{1}$ or $\textbf{k}=-\textbf{K}_{1}$,
which is the origin of
the strong enhancement of $\frac{dP_{x}}{d\theta}$.
However, since the systems becomes gapless at this point, leakage currents
will flow in reality, which disturbs the charge polarization.

\begin{figure}[t]
\centering
\includegraphics[width=8.5 cm]{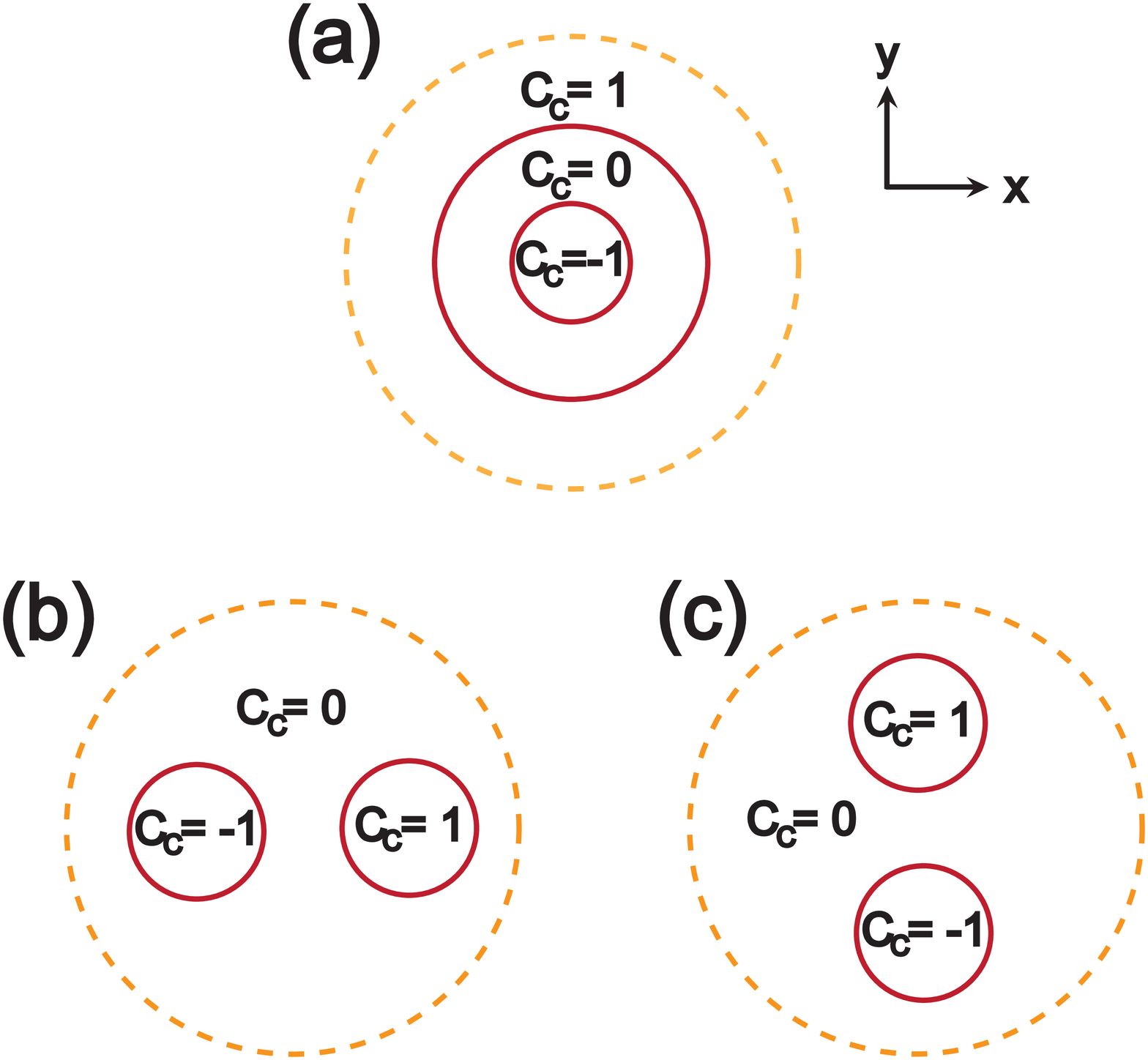}
\caption{(Color online)
The distribution of local charge Chern numbers within a skyrmion
in the case of $(t_{2}-\mu_{S})<m_{\text{AF}}<(t_{2}+\mu_{S})$ and $t_{2}>\mu_{\text{S}}>0$.
Here the skyrmion configurations for different orientations of $\textbf{m}_{\text{AF}}$
relative to $\textbf{t}_{2}=t_{2}\hat{z}$ are considered separately.
The largest dotted circle describes the size of a skyrmion.
Red solid lines indicate the phase boundary where the charge Chern number changes
due to gap-closing.
(a) For $\textbf{m}_{\text{AF}}=m_{\text{AF}}\hat{z}$ $(\textbf{t}_{2}\parallel\textbf{m}_{\text{AF}})$.
(b) For $\textbf{m}_{\text{AF}}=m_{\text{AF}}\hat{x}$ $(\textbf{t}_{2}\perp\textbf{m}_{\text{AF}})$.
(c) For $\textbf{m}_{\text{AF}}=m_{\text{AF}}\hat{y}$ $(\textbf{t}_{2}\perp\textbf{m}_{\text{AF}})$.
}\label{fig:localChern}
\end{figure}

The properties of skyrmions in CMI satisfying $|t_{2}-\mu_{S}|<|m_{\text{AF}}|<|t_{2}+\mu_{S}|$
can also be extracted from
the spatial distribution of the local charge Chern number $C_{C}(\textbf{r},t)=C_{C}(x,y)$.
We consider the following simple skyrmion configuration,
\begin{align}\label{eqn:skyrmionExample}
\hat{N}(x,y)=(\frac{2x}{x^{2}+y^{2}+1},\frac{2y}{x^{2}+y^{2}+1},\frac{x^{2}+y^{2}-1}{x^{2}+y^{2}+1}),
\end{align}
which corresponds to the case of $\textbf{m}_{\text{AF}} \parallel\textbf{t}_{2}$ for $\textbf{t}_{2}=t_{2}\hat{z}$.
For $(t_{2}-\mu_{S})<m_{\text{AF}}<(t_{2}+\mu_{S})$, the local charge Chern number has
spatial variation whose distribution is shown in Fig.~\ref{fig:localChern} (a).
The red solid circle connects the points where the bulk gap vanishes locally and the Chern number jumps.
Due to the occurrence of the gap-closing, the skyrmion defect cannot be constructed
from the adiabatic deformation of the spin ordering directions and
it is not energetically favorable to make skyrmions in this phase.

Because of the spin anisotropy induced by $\textbf{t}_{2}$,
the distribution of the local charge Chern numbers also depends
on the orientation of $\textbf{m}_{\text{AF}}$.
The local Chern number distribution for $\textbf{t}_{2}\perp\textbf{m}_{\text{AF}}$,
which can be described by permuting the components of $\hat{N}(x,y)$ in Eq.~(\ref{eqn:skyrmionExample}),
are shown in Fig.~\ref{fig:localChern} (b) and (c).
The rotation of $\textbf{m}_{\text{AF}}$ relative to $\textbf{t}_{2}$
changes the location of the circles.
The size of the solid circle depends on the magnitude of $m_{\text{AF}}$.
The topological equivalence of the distribution of the local Chern numbers
in Fig.~\ref{fig:localChern} (a), (b), and (c) can be seen
after compactifying the boundary of the skyrmion (the dotted circle) into a single point.
Therefore a skyrmion configuration always accompanies the band gap-closing.

\section{\label{sec:Conclusion} Conclusion}
In this work, we have investigated the adiabatic charge and spin polarizations
in chiral ferrimagnetic insulators through the smooth tilting of the inhomogeneous spin structures.
There are two essential ingredients for the adiabatic polarizations.
One is the bulk topological property and the other is the spin anisotropy.
In the case of CMI derived from AHI, the charge and spin polarizations
are solely determined by the bulk topological properties represented
by the charge and spin Chern numbers.
The charge and spin Hall effects induced by the spin gauge flux, which can be interpreted
as dual responses compared to the usual Hall effects, drive
the charge and spin polarizations.
Because of its topological nature, quantized pumping is possible
and a skyrmion carries quantum numbers in this CMI.
In particular, when the system is lack of the spatial inversion symmetry,
a new CMI phase dubbed the topological chiral magnetic insulator (TCMI) exists.
In TCMIs, the quantized charge and spin pumpings are possible
and the skyrmion texture is a spin-1/2 fermion with the charge $e$.

On the other hand, spin anisotropy plays
the crucial role in the case of the charge polarization in CMIs
derived from SHI.
Because of the spin anisotropy induced by the spin-orbit coupling,
the charge polarization occurs even in the topologically trivial SDW phases.
The charge polarization is also strongly affected by the bulk topological property.
When the magnitude of the spin ordering becomes comparable to
the spin-orbit coupling strength which corresponds to
TCMI phase in the collinear limit, the local charge Chern number
changes discontinuously in the position/time spaces.
Approaching the local phase boundary where the discontinuity in the local charge Chern number occurs,
the charge polarization shows
a strong enhancement due to the existence of gap-closing points.
In addition, in the case of a meron type defect, it can be charged in contrast to
the case of skyrmions. In particular, the meron
of the easy-plane Neel state carries finite charge
but the net charge is not quantized.

The quantum number of a skyrmion, which exists in CMI derived from TBI (both AHI and SHI) can be summarized
in the following way.
If the collinear magnetic ground state has finite Chern numbers independent
of the spatial orientation of the spins and
supports a stable skyrmion defect, the skyrmion carries a nonzero quantum number.
In the case of CMI derived from AHI, the relation of the skyrmion quantum number
with Chern numbers of the collinear magnetic ground state is obvious.
The charge quantum number of a skyrmion is given by
the spin Chern number.
On the other hand, the spin quantum number and quantum statistics of a skyrmion
are determined by the charge Chern number.
In the case of CMI derived from SHI,
since the spin Chern number is not well-defined, the topological
property of CMI is determined by the charge Chern number.
In both SDW and SHI phases, since the charge Chern number is zero,
there is no topological contribution to the skyrmion quantum numbers.
On the other hand, in the case of TCMI, although it has a finite charge Chern number,
a skyrmion is not a stable object because it always accompanies gap-closing.
Therefore the stable skyrmion texture of CMI derived from SHI does not carry nonzero quantum numbers,
which is consistent with the charge Chern number of the collinear magnetic ground state.


To observe the inhomogeneity induced adiabatic responses,
it is necessary to search systems in which both electron correlations and spin-orbit coupling
are essential ingredients determining the material properties.
Especially, to realize the TCMI, it is required to break lattice inversion symmetry.
In this regard, the bilayer of perovskite-type transition-metal oxides
grown along the [111] crystallographic axis
is a promising playground to realize the TCMI phase
because the strength of the inversion symmetry breaking potential can be easily manipulated
in this system.
The CMI with inhomogeneous spin structures
would be a fascinating venue to observe the interplay between
the magnetism and electricity coupled through the nontrivial topological invariants.~\cite{Shindou}


This work is supported by the Japan Society for the Promotion
of Science (JSPS) through the ``Funding Program for World-Leading Innovative
R$\&$D on Science and Technology (FIRST Program)".


\appendix

\section{\label{sec:GradientExpansion} Adiabatic polarization current from Moyal product expansion}
In this Appendix, we explicitly demonstrate how the adiabatic polarization
currents of homogeneous/inhomogeneous systems can be derived
using the gradient expansion (or Moyal product expansion) method.
For the derivation, we have referred the discussions in Ref.~\onlinecite{GE_Rammer, GE_Gurarie, GE_Baraff, Volovik_Book}.

\subsection{Polarization current in a system with translational invariance}
For a translationally invariant system,
the total charge current along the $\alpha$-direction ($\alpha=x, y$) can be written as
\begin{align}
J_{\alpha}=ie\int\frac{d^{2}p d\omega}{(2\pi\hbar)^3}\text{tr}[G^{-1}(\textbf{p},\omega)\partial_{p_{\alpha}}G(\textbf{p},\omega)],
\end{align}
where the symbol $\text{tr}$ denotes the summation over all discrete indices.
Throughout this section, we keep the unit $\hbar$ because
the gradient expansion is basically equivalent to the semi-classical
expansion in powers of $\hbar$.
Let us first consider the adiabatic current induced by the smooth
temporal variation of the system.
We assume that the system maintains the translational invariance
and the time dependence of the Hamiltonian appears
due to the coupling of electrons to the time dependent the order parameter field $\hat{N}(t)$.
Then the Green's function is, in general, a matrix possessing off-diagonal components
in the time space such as $G=G(\textbf{p};t,t')\equiv\langle t|\hat{G}(\textbf{p})|t'\rangle$.
The charge current can also be written as
\begin{align}\label{eqn:current}
J_{\alpha}=ie\int\frac{d^{2}p}{(2\pi\hbar)^2}\int dt\text{tr}\langle t|\hat{G}^{-1}(\textbf{p})\partial_{p_{\alpha}}\hat{G}(\textbf{p})|t\rangle.
\end{align}

If the order parameter field $\hat{N}(t)$ changes slowly in time,
the time dependence of the Green's function can be treated by using
the Wigner transformation.
In general, the Wigner transformation of a function
$F(t_{1},\textbf{x}_{1};t_{2},\textbf{x}_{2})\equiv F(x_{1};x_{2})$
is defined as
\begin{align}
\widetilde{F}(p,R)=\int dr e^{-\frac{i}{\hbar}p\cdot r}
F(R+r/2,R-r/2),
\end{align}
where the center-of-mass coordinate $R=\frac{x_{1}+x_{2}}{2}$ and the relative coordinate $r=x_{1}-x_{2}$
with $x_{1,2}=(t_{1,2},\textbf{x}_{1,2})$. Also,
$p=(\omega,\textbf{p})$, $p\cdot r = \textbf{p}\cdot \textbf{x} - \omega t$.
If $F$ is a convolution of two functions such as
\begin{align}
F(x,z)=\int dy A(x,y)B(y,z).
\end{align}
Then the Wigner transformations of $F$, $A$, $B$, which are represented by $\widetilde{F}$, $\widetilde{A}$, $\widetilde{B}$, respectively,
satisfy the following relation,
\begin{align}\label{eqn:gradientexpansion}
\widetilde{F}(p,x)=&\widetilde{A}(p,x)\widetilde{B}(p,x)
\nonumber\\
-&\frac{i\hbar}{2}[(\partial_{p_{i}} \widetilde{A})(\partial_{x_{i}} \widetilde{B})-(\partial_{x_{i}} \widetilde{A})(\partial_{p_{i}} \widetilde{B})]
\nonumber\\
-&\frac{\hbar^{2}}{8}[(\partial_{p_{i}}\partial_{p_{j}} \widetilde{A})(\partial_{x_{i}}\partial_{x_{j}} \widetilde{B})
+(\partial_{x_{i}}\partial_{x_{j}} \widetilde{A})(\partial_{p_{i}}\partial_{p_{j}} \widetilde{B})
\nonumber\\
&\quad-2(\partial_{p_{i}}\partial_{x_{j}} \widetilde{A})(\partial_{x_{i}}\partial_{p_{j}} \widetilde{B})
] + O(\hbar^{3}).
\end{align}
It shows that $\widetilde{F}$ can be written as a series of the space-time gradients of $\widetilde{A}$ and $\widetilde{B}$
in which the expansion parameter is given by $\hbar(\frac{\partial}{\partial_{x_{i}}})(\frac{\partial}{\partial_{p_{i}}})$.
In addition, the trace of $F$ satisfies
\begin{align}\label{eqn:trace}
\text{Tr}[F]=\int dx F(x,x)=\int \frac{dxdp}{(2\pi\hbar)^d} \widetilde{A}(p,x)\widetilde{B}(p,x),
\end{align}
where $d$ is the space-time dimension.

Using the relation in Eq.~(\ref{eqn:trace}), the total charge current $J_{\alpha}$
(Eq.~(\ref{eqn:current})) is given by
\begin{align}\label{eqn:currentfirst}
J_{\alpha}=-ie\int\frac{d^{2}p}{(2\pi\hbar)^2}\int \frac{dtd\omega}{(2\pi\hbar)}
\text{tr}[\widetilde{G}(\textbf{p},\omega,t)\partial_{p_{\alpha}}\widetilde{K}(\textbf{p},\omega,t)],
\end{align}
where $\widetilde{K}$ is the Wigner transformation of $K \equiv G^{-1}$ and
$\widetilde{G}$ is the Wigner transformed Green's function that is defined as
\begin{align}\label{eqn:WTGreenftn}
\widetilde{G}(\textbf{p},\omega,t)=\int d(t_{1}-t_{2}) e^{\frac{i}{\hbar}\omega(t_{1}-t_{2})} G(\textbf{p},t_{1},t_{2}),
\end{align}
where $t=(t_{1}+t_{2})/2$.

For the Hamiltonian $\hat{H}$ which is given by
\begin{align}
\hat{H}=\int\frac{d\textbf{p}}{(2\pi\hbar)^2}\psi^{\dag}(\textbf{p},t)H_{0}(\textbf{p},t)\psi(\textbf{p},t),
\end{align}
$G$ and $K=G^{-1}$ satisfy the
following relations,
\begin{align}
\Big[i\hbar \frac{\partial}{\partial t_{1}}-H_{0}(\textbf{p},t_{1})\Big]G(\textbf{p},t_{1},t_{2})=\hbar \delta(t_{1}-t_{2}),
\end{align}
and
\begin{align}\label{eqn:KG}
\int dt_{3} K(\textbf{p},t_{1},t_{3})G(\textbf{p},t_{3},t_{2})=\hbar \delta(t_{1}-t_{2}).
\end{align}
Therefore $K$ is given by
\begin{align}
K(\textbf{p},t_{1},t_{2})&=\Big[i\hbar \frac{\partial}{\partial t_{1}}-H_{0}(\textbf{p},t_{1})\Big]\delta(t_{1}-t_{2}),
\nonumber\\
&=\int\frac{d\Omega}{2\pi\hbar}\Big[\Omega-H_{0}(\textbf{p},\frac{t_{1}+t_{2}}{2})\Big]e^{-\frac{i}{\hbar}\Omega(t_{1}-t_{2})}.
\end{align}
After the Wigner transformation of $K$, we obtain,
\begin{align}
\widetilde{K}(\textbf{p},\omega,t)=\omega-H_{0}(\textbf{p},t)\equiv \mathcal{G}^{-1}(\textbf{p},\omega,t),
\end{align}
where $\mathcal{G}$ is the semi-classical Green's function in which $\omega$ and $t$
are treated as independent variables.

The next step is to express the Wigner transformed Green's function $\widetilde{G}$
as a function of the semi-classical Green's function $\mathcal{G}$.
Let us first perform the Wigner transformation of Eq.~(\ref{eqn:KG}).
From Eq.~(\ref{eqn:gradientexpansion}), we expect that
the Wigner transformation of the left hand side of Eq.~(\ref{eqn:KG})
can be expanded in powers of $\hbar$, which leads to following relation,
\begin{align}\label{eqn:Thetaseries}
\Theta[\widetilde{K},\widetilde{G}]\equiv\sum_{j=0}^{\infty}\hbar^{j}\Theta_{j}[\widetilde{K},\widetilde{G}]=\hbar,
\end{align}
in which
\begin{align}
\Theta_{0}[\widetilde{K},\widetilde{G}]&=\widetilde{K}\widetilde{G},
\nonumber\\
\Theta_{1}[\widetilde{K},\widetilde{G}]&=
-\frac{i}{2}[(\partial_{p_{i}} \widetilde{K})(\partial_{x_{i}} \widetilde{G})-(\partial_{x_{i}} \widetilde{K})(\partial_{p_{i}} \widetilde{G})],
\nonumber\\
\Theta_{2}[\widetilde{K},\widetilde{G}]&=
-\frac{1}{8}[(\partial_{p_{i}}\partial_{p_{j}} \widetilde{K})(\partial_{x_{i}}\partial_{x_{j}} \widetilde{G})
+(\partial_{x_{i}}\partial_{x_{j}} \widetilde{K})(\partial_{p_{i}}\partial_{p_{j}} \widetilde{G})
\nonumber\\
&\qquad-2(\partial_{p_{i}}\partial_{x_{j}} \widetilde{K})(\partial_{x_{i}}\partial_{p_{j}} \widetilde{G})
] ,
\end{align}
and so on. Now we make expansions of $\widetilde{G}$ and $\widetilde{K}$ in powers of $\hbar$
in such a way as,
\begin{align}\label{eqn:GKseries}
\widetilde{G}=\sum_{k=0}^{\infty}\hbar^{k+3}\widetilde{G}_{k},\quad
\widetilde{K}=\sum_{l=0}^{\infty}\hbar^{l}\widetilde{K}_{l},
\end{align}
Putting Eq.~(\ref{eqn:GKseries}) into Eq.~(\ref{eqn:Thetaseries}), we obtain
\begin{align}
\Theta_{0}[\widetilde{K}_{0},\hbar^{3}\widetilde{G}_{0}]=\hbar,
\end{align}
and
\begin{align}
\sum_{j+k+l>0}\Theta_{j}[\widetilde{K}_{l},\widetilde{G}_{k}]=0,
\end{align}
where the summation is performed for $j,k,l$ for given $j+k+l>0$.
Taking into account of the fact that $\widetilde{K}=\widetilde{K}_{0}=\mathcal{G}^{-1}$,
$\widetilde{G}_{0,1,2}$ can be obtained as
\begin{align}\label{eqn:WTGseries}
\widetilde{G}_{0}&=\frac{1}{\hbar^{2}}\mathcal{G},
\nonumber\\
\widetilde{G}_{1}&=
\frac{i}{2\hbar^{2}}\mathcal{G}[(\partial_{p_{i}} \mathcal{G}^{-1})(\partial_{x_{i}} \mathcal{G})
-(\partial_{x_{i}} \mathcal{G}^{-1})(\partial_{p_{i}} \mathcal{G})],
\nonumber\\
\widetilde{G}_{2}&=
\frac{1}{8\hbar^{2}}\mathcal{G}[(\partial_{p_{i}}\partial_{p_{j}} \mathcal{G}^{-1})(\partial_{x_{i}}\partial_{x_{j}} \mathcal{G})
+(\partial_{x_{i}}\partial_{x_{j}} \mathcal{G}^{-1})(\partial_{p_{i}}\partial_{p_{j}} \mathcal{G})
\nonumber\\
&\quad\qquad-2(\partial_{p_{i}}\partial_{x_{j}} \mathcal{G}^{-1})(\partial_{x_{i}}\partial_{p_{j}} \mathcal{G})
]
\nonumber\\
&\quad+\frac{1}{4\hbar^{2}}\mathcal{G}(\partial_{x_{j}} \mathcal{G}^{-1})(\partial_{p_{j}}[\mathcal{G}(\partial_{p_{i}}\mathcal{G}^{-1} )(\partial_{x_{i}} \mathcal{G})])
\nonumber\\
&\quad-\frac{1}{4\hbar^{2}}\mathcal{G}(\partial_{x_{j}} \mathcal{G}^{-1})(\partial_{p_{j}}[\mathcal{G}(\partial_{x_{i}}\mathcal{G}^{-1} )(\partial_{p_{i}} \mathcal{G})])
\nonumber\\
&\quad-\frac{1}{4\hbar^{2}}\mathcal{G}(\partial_{p_{j}} \mathcal{G}^{-1})(\partial_{x_{j}}[\mathcal{G}(\partial_{p_{i}}\mathcal{G}^{-1} )(\partial_{x_{i}} \mathcal{G})])
\nonumber\\
&\quad+\frac{1}{4\hbar^{2}}\mathcal{G}(\partial_{p_{j}} \mathcal{G}^{-1})(\partial_{x_{j}}[\mathcal{G}(\partial_{x_{i}}\mathcal{G}^{-1} )(\partial_{p_{i}} \mathcal{G})]).
\end{align}

Considering a smooth variation of the order parameter field $\hat{N}(t)$ in time,
the leading order contribution to $\widetilde{G}$ is given by
\begin{align}\label{eqn:WTGfirst}
\widetilde{G}=&\hbar\mathcal{G}
-\frac{i\hbar^{2}}{2}\mathcal{G}[(\partial_{\omega}\mathcal{G}^{-1} )(\partial_{t} \mathcal{G})
-(\partial_{t} \mathcal{G}^{-1})(\partial_{\omega} \mathcal{G})].
\end{align}
Therefore, from Eq.~(\ref{eqn:currentfirst}) and Eq.~(\ref{eqn:WTGfirst}), the polarization current induced by
the adiabatic temporal variation of the order parameter field $\hat{N}(t)$ is given by
\begin{align}
J^{(1)}_{\alpha}
=&-e\hbar^{2}\int\frac{d^{2}p}{(2\pi\hbar)^2}\int \frac{dtd\omega}{(4\pi\hbar)}
\text{tr}[\mathcal{G}\{(\partial_{\omega}\mathcal{G}^{-1} )(\partial_{t} \mathcal{G})
\nonumber\\
&\qquad-(\partial_{t} \mathcal{G}^{-1})(\partial_{\omega} \mathcal{G})\}(\partial_{p_{\alpha}}\mathcal{G}^{-1})],
\nonumber\\
=&e\hbar^{2}\int\frac{d^{2}p}{(2\pi\hbar)^2}\int \frac{dtd\omega}{(4\pi\hbar)}
\text{tr}[\{(\partial_{\omega}\mathcal{G}^{-1} )\mathcal{G}(\partial_{t} \mathcal{G}^{-1})\mathcal{G}
\nonumber\\
&\qquad-(\partial_{t} \mathcal{G}^{-1})\mathcal{G}(\partial_{\omega} \mathcal{G}^{-1})\mathcal{G}\}
(\partial_{p_{\alpha}}\mathcal{G}^{-1})\mathcal{G}],
\nonumber\\
=&e\hbar^{2}\int\frac{d^{2}p}{(2\pi\hbar)^2}\int \frac{dtd\omega}{(12\pi\hbar)}
\nonumber\\
&\qquad\times\varepsilon_{\mu\nu\lambda}\text{tr}[(\partial_{\mu}\mathcal{G}^{-1} )\mathcal{G}(\partial_{\nu} \mathcal{G}^{-1})\mathcal{G}
(\partial_{\lambda}\mathcal{G}^{-1})\mathcal{G}],
\end{align}
where $(\mu,\nu,\lambda)$ run over $(\omega,t,p_{\alpha})$.
In particular, considering a cyclic variation of $\hat{N}(t)=\hat{N}(\theta(t))$,
which can be parameterized using a polar angle $\theta(t)\in[0,2\pi]$,
the adiabatic current along the $x$-direction $J^{(1)}_{x}$, for example, can be written as
\begin{align}
J^{(1)}_{x}
=&e\int\frac{dp_{y}}{2\pi\hbar} \Big\{\int \frac{dp_{x}d\theta d\omega}{24\pi^{2}}
\nonumber\\
&\qquad\times\varepsilon_{\mu\nu\lambda}\text{tr}[(\partial_{\mu}\mathcal{G}^{-1} )\mathcal{G}(\partial_{\nu} \mathcal{G}^{-1})\mathcal{G}
(\partial_{\lambda}\mathcal{G}^{-1})\mathcal{G}]\Big\},
\end{align}
where $(\mu,\nu,\lambda)=(\omega,\theta,p_{x})$. It is interesting
to notice that the expression in the large parenthesis $\{\}$
is nothing but the first Chern number, which is the topological
origin for the quantized charge pumping through the cyclic variation
of the order parameter field $\hat{N}(t)$.
The electric charge induced by the one dimensional smooth structure
can be described by the same topological invariant if $t$ is replaced by the spatial coordinate $x$,
as discussed in the recent work by V\"{a}yrynen et al..~\cite{Vayrynen}

\subsection{Polarization current in a system with spatial inhomogeneity}
Now we consider the adiabatic polarization current
in inhomogeneous systems.
In particular, we focus on the contribution of
the spatial inhomogeneity to the adiabatic current.
For simplicity, we assume that the system maintains
the translational invariance along the $x$-direction and the spatial modulation occurs
along the $y$-direction. Using the Wigner transformation, the total charge current
of the system can be written as
\begin{align}\label{eqn:2ndcurrent}
J_{x}&=-ie\int\frac{dp_{x}}{2\pi\hbar}\int dy\int dt\text{tr}\langle t,y|\hat{G}(p_{x})\partial_{p_{x}}\hat{K}(p_{x})|t,y\rangle,
\nonumber\\
&=-ie\int\frac{dp_{x}}{2\pi\hbar}\int \frac{dydp_{y}}{2\pi\hbar}\int \frac{dtd\omega}{2\pi\hbar}
\nonumber\\
&\qquad\times\text{tr}\big[\widetilde{G}(p_{x};p_{y},y;\omega,t)\partial_{p_{x}}\widetilde{K}(p_{x};p_{y},y;\omega,t)\big].
\end{align}
Smooth variation in the $(t,y)$ space can be treated by using
the gradient expansion method following the similar procedures
taken in the previous subsection.

We consider a class of Hamiltonians which can be written as
\begin{align}
\hat{H}=\int\frac{dp_{x}}{(2\pi\hbar)}\int dy\psi^{\dag}(p_{x},y,t)H_{0}(p_{x},\frac{\hbar}{i}\frac{\partial}{\partial_{y}},y,t)\psi(p_{x},y,t),
\end{align}
in which $H_{0}$ has the dependence on both $y$ and $\frac{\partial}{\partial_{y}}$.
The inverse Green's function $K$ is given by
\begin{align}\label{eqn:SCGreenftn}
&K(p_{x};y_{1},t_{1};y_{2},t_{2})
\nonumber\\
&=\Big[i\hbar \frac{\partial}{\partial t_{1}}-H_{0}(p_{x},\frac{\hbar}{i}\frac{\partial}{\partial_{y_{1}}},y_{1},t_{1})\Big]\delta(t_{1}-t_{2})\delta(y_{1}-y_{2}),
\nonumber\\
&=\int\frac{d\Omega dp_{y}}{(2\pi\hbar)^{2}}
\Big[\Omega-H_{0}(p_{x},p_{y},\frac{y_{1}+y_{2}}{2},\frac{t_{1}+t_{2}}{2})\Big]
\nonumber\\
&\qquad\times e^{\frac{i}{\hbar}[p_{y}(y_{1}-y_{2})-\Omega(t_{1}-t_{2})]}.
\end{align}
After the Wigner transformation, $\widetilde{K}$ is given by
\begin{align}
\widetilde{K}(p_{x};p_{y},y;\omega,t)=\omega-H_{0}(p_{x},p_{y},y,t)\equiv \mathcal{G}^{-1},
\end{align}
where $\mathcal{G}$ is the semi-classical Green's function in which
$(\omega,t)$ and $(p_{y},y)$ are treated as independent variables.

To describe the inhomogeneity induced adiabatic current, we consider
the gradient expansion of $\widetilde{G}$ in powers of $\hbar$.
Since we are interested in the leading order
contribution of the spatial inhomogeneity to the adiabatic polarization current,
only the terms which are linear in the derivatives of $t$ and $y$, respectively, are important.
These are included in the $\widetilde{G}_{2}$ in Eq.~(\ref{eqn:WTGseries}).
The expression of the adiabatic current can be simplified by considering
the following relations. At first,
$\partial_{p_{i}}\partial_{p_{j}} \mathcal{G}^{-1}=\partial_{\omega}\partial_{p_{y}} \mathcal{G}^{-1}=0$
from the general structure of the $\mathcal{G}$ in Eq.~(\ref{eqn:SCGreenftn}).
Similarly, $\partial_{\omega}\partial_{t} \mathcal{G}^{-1}=\partial_{\omega}\partial_{y} \mathcal{G}^{-1}=0$.
Then the leading order contribution to the inhomogeneity induced polarization current can
be summarized as
\begin{align}
J^{(2)}_{x}=J^{(2,A)}_{x}+J^{(2,B)}_{x}+J^{(2,C)}_{x}+J^{(2,D)}_{x},
\end{align}
in which
\begin{align}\label{eqn:J2A}
J^{(2,A)}_{x}
&=-\frac{ie\hbar^{3}}{4}\int\frac{dp_{x}}{2\pi\hbar}\int \frac{dydp_{y}}{2\pi\hbar}\int \frac{dtd\omega}{2\pi\hbar}
\nonumber\\
&
\times\text{tr}\big[\mathcal{G}(\partial_{t}\partial_{y}\mathcal{G}^{-1} )\mathcal{G}
\big\{
(\partial_{\omega} \mathcal{G}^{-1})\mathcal{G}(\partial_{p_{y}} \mathcal{G}^{-1})\mathcal{G}(\partial_{p_{x}} \mathcal{G}^{-1})
\nonumber\\
&\qquad\qquad\qquad\quad+(\partial_{p_{y}} \mathcal{G}^{-1})\mathcal{G}(\partial_{\omega} \mathcal{G}^{-1})\mathcal{G}(\partial_{p_{x}} \mathcal{G}^{-1})
\nonumber\\
&\qquad\qquad\qquad\quad+(\partial_{p_{x}} \mathcal{G}^{-1})\mathcal{G}(\partial_{\omega} \mathcal{G}^{-1})\mathcal{G}(\partial_{p_{y}} \mathcal{G}^{-1})
\nonumber\\
&\qquad\qquad\qquad\quad+(\partial_{p_{x}} \mathcal{G}^{-1})\mathcal{G}(\partial_{p_{y}} \mathcal{G}^{-1})\mathcal{G}(\partial_{\omega} \mathcal{G}^{-1})
\nonumber\\
&\qquad\qquad\qquad\quad-(\partial_{\omega} \mathcal{G}^{-1})\mathcal{G}(\partial_{p_{x}} \mathcal{G}^{-1})\mathcal{G}(\partial_{p_{y}} \mathcal{G}^{-1})
\nonumber\\
&\qquad\qquad\qquad\quad-(\partial_{p_{y}} \mathcal{G}^{-1})\mathcal{G}(\partial_{p_{x}} \mathcal{G}^{-1})\mathcal{G}(\partial_{\omega} \mathcal{G}^{-1})
\big\}\big],
\end{align}
and
\begin{align}\label{eqn:J2B}
J^{(2,B)}_{x}
&=\frac{ie\hbar^{3}}{4}\int\frac{dp_{x}}{2\pi\hbar}\int \frac{dydp_{y}}{2\pi\hbar}\int \frac{dtd\omega}{2\pi\hbar}
\nonumber\\
&
\times\text{tr}\big[\mathcal{G}(\partial_{t}\partial_{p_{y}}\mathcal{G}^{-1} )\mathcal{G}
\big\{
(\partial_{\omega} \mathcal{G}^{-1})\mathcal{G}(\partial_{y} \mathcal{G}^{-1})\mathcal{G}(\partial_{p_{x}} \mathcal{G}^{-1})
\nonumber\\
&\qquad\qquad\qquad\quad+(\partial_{y} \mathcal{G}^{-1})\mathcal{G}(\partial_{\omega} \mathcal{G}^{-1})\mathcal{G}(\partial_{p_{x}} \mathcal{G}^{-1})
\nonumber\\
&\qquad\qquad\qquad\quad+(\partial_{p_{x}} \mathcal{G}^{-1})\mathcal{G}(\partial_{\omega} \mathcal{G}^{-1})\mathcal{G}(\partial_{y} \mathcal{G}^{-1})
\nonumber\\
&\qquad\qquad\qquad\quad+(\partial_{p_{x}} \mathcal{G}^{-1})\mathcal{G}(\partial_{y} \mathcal{G}^{-1})\mathcal{G}(\partial_{\omega} \mathcal{G}^{-1})
\nonumber\\
&\qquad\qquad\qquad\quad-(\partial_{\omega} \mathcal{G}^{-1})\mathcal{G}(\partial_{p_{x}} \mathcal{G}^{-1})\mathcal{G}(\partial_{y} \mathcal{G}^{-1})
\nonumber\\
&\qquad\qquad\qquad\quad-(\partial_{y} \mathcal{G}^{-1})\mathcal{G}(\partial_{p_{x}} \mathcal{G}^{-1})\mathcal{G}(\partial_{\omega} \mathcal{G}^{-1})
\big\}\big],
\end{align}
and
\begin{align}\label{eqn:J2C}
J^{(2,C)}_{x}
=&-\frac{ie\hbar^{3}}{4}\int\frac{dp_{x}}{2\pi\hbar}\int \frac{dydp_{y}}{2\pi\hbar}\int \frac{dtd\omega}{2\pi\hbar}
\nonumber\\
&\times\Big\{\text{tr}\big[\mathcal{G}(\partial_{p_{i}}\mathcal{G}^{-1} )\mathcal{G}
\{(\partial_{p_{j}} \mathcal{G}^{-1})\mathcal{G}(\partial_{x_{i}} \mathcal{G}^{-1})
\nonumber\\
&\qquad+(\partial_{x_{i}} \mathcal{G}^{-1})\mathcal{G}(\partial_{p_{j}} \mathcal{G}^{-1})\}\mathcal{G}(\partial_{p_{x}} \mathcal{G}^{-1})
\mathcal{G}(\partial_{x_{j}}\mathcal{G}^{-1})]
\nonumber\\
&\quad+\text{tr}[\mathcal{G}(\partial_{x_{i}}\mathcal{G}^{-1} )\mathcal{G}
\{(\partial_{x_{j}} \mathcal{G}^{-1})\mathcal{G}(\partial_{p_{i}} \mathcal{G}^{-1})
\nonumber\\
&\qquad+(\partial_{p_{i}} \mathcal{G}^{-1})\mathcal{G}(\partial_{x_{j}} \mathcal{G}^{-1})\}\mathcal{G}(\partial_{p_{x}} \mathcal{G}^{-1})
\mathcal{G}(\partial_{p_{j}}\mathcal{G}^{-1})]
\nonumber\\
&\quad-\text{tr}[\mathcal{G}(\partial_{x_{i}}\mathcal{G}^{-1} )\mathcal{G}
\{(\partial_{p_{i}} \mathcal{G}^{-1})\mathcal{G}(\partial_{p_{j}} \mathcal{G}^{-1})
\nonumber\\
&\qquad+(\partial_{p_{j}} \mathcal{G}^{-1})\mathcal{G}(\partial_{p_{i}} \mathcal{G}^{-1})\}\mathcal{G}(\partial_{p_{x}} \mathcal{G}^{-1})
\mathcal{G}(\partial_{x_{j}}\mathcal{G}^{-1})]
\nonumber\\
&\quad-\text{tr}[\mathcal{G}(\partial_{p_{i}}\mathcal{G}^{-1} )\mathcal{G}
\{(\partial_{x_{i}} \mathcal{G}^{-1})\mathcal{G}(\partial_{x_{j}} \mathcal{G}^{-1})
\nonumber\\
&\qquad+(\partial_{x_{j}} \mathcal{G}^{-1})\mathcal{G}(\partial_{x_{i}} \mathcal{G}^{-1})\}\mathcal{G}(\partial_{p_{x}} \mathcal{G}^{-1})
\mathcal{G}(\partial_{p_{j}}\mathcal{G}^{-1})\big]\Big\},
\end{align}
and
\begin{align}\label{eqn:J2D}
J^{(2,D)}_{x}
=&-\frac{ie\hbar^{3}}{4}\int\frac{dp_{x}}{2\pi\hbar}\int \frac{dydp_{y}}{2\pi\hbar}\int \frac{dtd\omega}{2\pi\hbar}
\nonumber\\
&\times\Big\{\text{tr}\big[\mathcal{G}(\partial_{p_{x}}\mathcal{G}^{-1} )\mathcal{G}
\{(\partial_{x_{i}} \mathcal{G}^{-1})\mathcal{G}(\partial_{p_{i}} \mathcal{G}^{-1})
\nonumber\\
&\qquad-(\partial_{p_{i}} \mathcal{G}^{-1})\mathcal{G}(\partial_{x_{i}} \mathcal{G}^{-1})\}\mathcal{G}(\partial_{p_{j}} \mathcal{G}^{-1})
\mathcal{G}(\partial_{x_{j}}\mathcal{G}^{-1})]
\nonumber\\
&\quad+\text{tr}[\mathcal{G}(\partial_{p_{x}}\mathcal{G}^{-1} )\mathcal{G}
\{(\partial_{p_{i}} \mathcal{G}^{-1})\mathcal{G}(\partial_{x_{i}} \mathcal{G}^{-1})
\nonumber\\
&\qquad-(\partial_{x_{i}} \mathcal{G}^{-1})\mathcal{G}(\partial_{p_{i}} \mathcal{G}^{-1})\}\mathcal{G}(\partial_{x_{j}} \mathcal{G}^{-1})
\mathcal{G}(\partial_{p_{j}}\mathcal{G}^{-1})\big]\Big\},
\end{align}
where $(x_{i},p_{i})=(t,\omega)$ or $(y,p_{y})$.
Here $J^{(2,A)}_{x}$ ($J^{(2,B)}_{x}$) contains the second derivative of $\mathcal{G}^{-1}$, i.e.,
$\partial_{t}\partial_{y}\mathcal{G}^{-1}$
($\partial_{t}\partial_{p_{y}}\mathcal{G}^{-1}$).
$J^{(2,C)}_{x}$ ($J^{(2,D)}_{x}$) consists of the terms which contains a symmetric (antisymmetric)
permutation of partial derivatives between parenthesis in the middle.

\underline{Lattice Dirac Hamiltonian:}
We apply the above result to the semi-classical lattice Dirac Hamiltonian,
which has the following structure,
\begin{align}
H_{\text{Dirac}}=\sum_{a=1}^{5}d_{a}(\textbf{p},\textbf{x})\Gamma_{a},
\end{align}
in which the five $4\times 4$ traceless Gamma matrices satisfy $\{\Gamma_{\mu},\Gamma_{\nu}\}=2\delta_{\mu,\nu}\mathbb{I}$
with the identity matrix $\mathbb{I}$.
$\vec{d}=(d_{1},d_{2},d_{3},d_{4},d_{5})$ is a five-component vector that depends
on the position $\textbf{x}$ and momentum $\textbf{p}$ coordinates.
Using the identity $\text{tr}[\Gamma_{i}\Gamma_{j}\Gamma_{k}\Gamma_{l}\Gamma_{m}]=-4\varepsilon_{ijklm}$,
it can be shown that
\begin{align}
J^{(2,A)}_{x}=J^{(2,B)}_{x}=0.
\end{align}
Regarding $J^{(2,C)}_{x}$ and $J^{(2,D)}_{x}$, it is crucial to
observe the following relation,
\begin{align}
&\int d\omega\text{tr}[\mathcal{G}(\partial_{\nu_{1}}\mathcal{G}^{-1})\cdot\cdot\cdot\mathcal{G}(\partial_{\nu_{5}}\mathcal{G}^{-1})]
\nonumber\\
&=\int d\omega\frac{1}{120}\varepsilon_{\nu_{1}\nu_{2}\nu_{3}\nu_{4}\nu_{5}}
\text{tr}[\mathcal{G}(\partial_{\nu_{1}}\mathcal{G}^{-1})\cdot\cdot\cdot\mathcal{G}(\partial_{\nu_{5}}\mathcal{G}^{-1})],
\end{align}
where the five indices $(\nu_{1},\nu_{2},\nu_{3},\nu_{4},\nu_{5})$ run over $(\omega,p_{x},p_{y},t,y)$.
The antisymmetry of the trace under the permutation of the indices $\nu_{i}$
leads to
\begin{align}
J^{(2,C)}_{x}&=0,
\end{align}
and
\begin{align}
J^{(2,D)}_{x}
=&-\frac{ie\hbar^{3}}{60}\int\frac{dp_{x}}{2\pi\hbar}\int \frac{dydp_{y}}{2\pi\hbar}\int \frac{dtd\omega}{2\pi\hbar}
\nonumber\\
&\times\varepsilon_{\nu_{1}\nu_{2}\nu_{3}\nu_{4}\nu_{5}}
\text{tr}[\mathcal{G}(\partial_{\nu_{1}}\mathcal{G}^{-1})\cdot\cdot\cdot\mathcal{G}(\partial_{\nu_{5}}\mathcal{G}^{-1})].
\end{align}

For the description of the charge pumping in topological
chiral magnets, we can assume that the spatial inhomogeneity of the system appears only through
the coupling to the inhomogeneous order parameter field $\hat{N}(y,t)$ and
the charge current is induced by the adiabatic variation of $\hat{N}(y,t)$,
which satisfies $\hat{N}^{2}=1$, i.e., $\hat{N}\in S^{2}$.
We introduce a spherical coordinate $(\theta,\phi)$ for the space-time $(y,t)$ dependent unit vector field $\hat{N}(y,t)=\hat{N}(\theta(y,t),\phi(y,t))$.
Then $J^{(2,D)}_{x}$ can be rewritten as
\begin{align}
J^{(2,D)}_{x}
=&-\frac{ie}{60}\int\frac{d\omega}{2\pi}\int_{T^{2}} \frac{d^{2}\textbf{p}}{(2\pi)^{2}}\int_{S^2} d\Omega(\theta,\phi)
\nonumber\\
&\times\varepsilon_{\mu_{1}\mu_{2}\mu_{3}\mu_{4}\mu_{5}}
\text{tr}[\mathcal{G}(\partial_{\mu_{1}}\mathcal{G}^{-1})\cdot\cdot\cdot\mathcal{G}(\partial_{\mu_{5}}\mathcal{G}^{-1})],
\nonumber\\
=&eC^{(2)}_{\text{Chern}},
\end{align}
where the indices $(\mu_{1},\mu_{2},\mu_{3},\mu_{4},\mu_{5})$ run over $(\omega,p_{x},p_{y},\theta,\phi)$.
Here $T^{2}$ indicates the 2-torus given by two dimensional Brillouin zone.
It is interesting to note that $C^{(2)}_{\text{Chern}}$ is nothing but the second Chern number,
which is quantized when the spherical solid angle $\Omega(\theta,\phi)$ is fully covered by $\hat{N}(y,t)$.
Therefore the inhomogeneity induced adiabatic current $J^{(2)}_{x}$ can be
summarized in the following suggestive way,
\begin{align}
J^{(2)}_{x}=J^{(2)}_{x,\text{topological}}+J^{(2)}_{x,\text{perturb}},
\end{align}
in which $J^{(2)}_{x,\text{topological}}=J^{(2,D)}_{x}$ is the topological
contribution to the adiabatic current whose topological nature is endowed with the second
Chern number.
All the other non-topological (perturbative) contributions are included in
$J^{(2)}_{x,\text{perturb}}=J^{(2,A)}_{x}+J^{(2,B)}_{x}+J^{(2,C)}_{x}$.




\begin{thebibliography}{99}
\bibitem{Footnote_SCGreen}
Compared to the low energy Hamiltonians for TCMIs in Eq.~(\ref{eqn:effH1}) and Eq.~(\ref{eqn:effH2}),
the inversion symmetry breaking term ($\mu_{S}$) is missing in this semi-classical Green's function.
However, as long as the the inversion symmetry breaking term does not induce
bulk band gap-closing, the topological property of the system can still
be described with given $\mathcal{G}$ here.
\end{thebibliography}
\end{document}